\documentclass{aa}  

\usepackage{siunitx}
\usepackage{enumitem}
\usepackage{gensymb}
\usepackage{graphicx}
\usepackage{lineno}
\usepackage{txfonts}
\usepackage{hyperref}
\usepackage{multicol}
\usepackage{capt-of}
\usepackage{ragged2e}
\hypersetup{
    colorlinks=true,
    urlcolor=blue,
    citecolor=blue,
    linkcolor=blue,
    }

\begin{document}

\title{Interstellar extinction of classical Cepheids \\
from neighboring stars\thanks{Based on observations collected at the European Southern Observatory under ESO programme 112.25ES.001.}}

   \subtitle{I. Methodology and application to 16 Galactic Cepheids}

   \author{Wilma Kiviaho
          \inst{1,2} \email{wilma.kiviaho@obspm.fr}
          \and
          Pierre Kervella\inst{1,2} \email{pierre.kervella@obspm.fr}
          \and
          Mika Juvela\inst{3} \email{mika.juvela@helsinki.fi}
          \and
          Alexandre Gallenne\inst{4} \email{agallenne@academicos.uta.cl}
          \and
          Nicolas Nardetto\inst{5} \email{nicolas.nardetto@oca.eu}
          \and
          Carine Babusiaux\inst{6} \email{carine.babusiaux@univ-grenoble-alpes.fr}
          \and
          Rosine Lallement\inst{1} \email{rosine.lallement@obspm.fr}
          \and
          Vincent Hocdé\inst{5} \email{vincent.hocde@oca.eu}
          \and
          Garance Bras\inst{1} \email{garance.bras@obspm.fr}
          \and
          Nancy Remage Evans\inst{7} \email{nevans@cfa.harvard.edu}
          \and
          Antoine Mérand\inst{8} \email{amerand@eso.org}
          \and
          Grzegorz Pietrzyński\inst{9} \email{pietrzyn@camk.edu.pl}
          \and
          Gergely Hajdu\inst{9} \email{ghajdu@camk.edu.pl}
          \and
          Henryka Netzel\inst{9} \email{henryka.netzel@gmail.com}
          \and
          Ricardo Salinas\inst{9} \email{rsalinas@camk.edu.pl}
          \and
          Bogumił Pilecki\inst{9} \email{pilecki@camk.edu.pl}
          \and
          Cezary Gałan\inst{9} \email{cgalan@camk.edu.pl}
          \and
          Frédéric Arenou\inst{10,1} \email{Frederic.Arenou@obspm.fr}
          \and
          Manon Bailleul\inst{5} \email{manon.bailleul@oca.eu}
          \and
          Katia Sivkova\inst{1,2} \email{ekaterina.sivkova@obspm.fr}
          \and
          Daniela González\inst{1,2} \email{daniela.gonzalez@obspm.fr}
          \and
          Louise Breuval\inst{11} \email{lbreuval@stsci.edu}
          \and
          Rajeev Singh Rathour\inst{5} \email{Rajeev.Rathour@oca.eu}
          \and
          Brankica Apostolova\inst{5} \email{brankica.apostolova@oca.eu}
          }

   \institute{LIRA, Observatoire de Paris, Université PSL, Sorbonne Université, Université Paris Cité, CY Cergy Paris Université, CNRS, 92190 Meudon, France
    \email{wilma.kiviaho@obspm.fr}
        \and
            French-Chilean Laboratory for Astronomy, IRL 3386, CNRS and U. de Chile, Casilla 36-D, Santiago, Chile
         \and
             Department of Physics, PO Box 64, 00014, University of Helsinki, Finland
         \and
             Instituto de Alta Investigación, Universidad de Tarapacá, Casilla 7D, Arica, Chile
         \and
             Université Côte d'Azur, Observatoire de la Côte d'Azur, CNRS, Laboratoire Lagrange, Nice, France
         \and
             Univ. Grenoble Alpes, CNRS, IPAG, 38000 Grenoble, France
         \and
             Smithsonian Astrophysical Observatory, MS 4, 60 Garden St., Cambridge, MA 02138, USA
         \and
             European Southern Observatory, Karl-Schwarzschild-Straße 2, 85748 Garching, Germany
         \and
             Nicolaus Copernicus Astronomical Centre, Polish Academy of Sciences, Bartycka 18, PL-00-716 Warszawa, Poland
         \and  
             UNIDIA, Observatoire de Paris, Université PSL, CNRS, 5 Place Jules Janssen, 92190, Meudon, France
         \and
             European Space Agency (ESA), ESA Office, Space Telescope Science Institute, 3700 San Martin Drive, Baltimore, MD 21218, USA
             }

  \abstract
  {Galactic Cepheid variable stars are fundamental calibrators of the cosmic distance scale through their Period–Luminosity (PL) relation. The high-precision parallaxes expected from Gaia Data Release 4 (DR4) will enable a calibration of the Milky Way PL relation with unprecedented precision. This will place stringent requirements on interstellar extinction corrections, which remain a dominant source of uncertainty.}
  {We strive to derive highly accurate extinction values for Galactic Cepheids. We present an innovative methodology relying on the analysis of neighboring field stars and apply it to a pilot sample of 16 Cepheids.}
  {We identify physically near field stars using Gaia DR3 astrometry. For these stars, we determine atmospheric parameters from medium- to high-resolution VLT/FLAMES spectroscopy. Fixing these parameters, we derive individual extinctions by fitting spectral energy distributions constructed from Gaia XP spectra and infrared photometry. We then reconstruct the three-dimensional distribution of extinction and interpolate extinction to the Cepheid position.}
  {We report extinction $A_{\mathrm V}$, total-to-selective extinction ratio $R_{\mathrm V}$, and color excess $E_{\mathrm B-V}$ for all 16 Cepheids. Our method achieves a mean precision of approximately $5\%$, improving upon typical literature values, with no significant systematic offset detected.}
  {We demonstrate that using field stars bypasses uncertainties associated with Cepheid variability and can provide robust extinction estimates for Galactic Cepheids.}

   \keywords{Stars: variables: Cepheids --
                dust, extinction --
                Stars: fundamental parameters --
                distance scale
               }

   \maketitle
   \nolinenumbers

\section{Introduction}
Cepheid variable stars are known for their empirical Period-Luminosity relation (PL), also called the Leavitt law \citep{Leavitt1912}, which connects their pulsation period with intrinsic brightness. This feature has granted Classical Cepheids (hereafter just Cepheids) a role as a fundamental calibrator in the extragalactic distance ladder. The fourth data release (DR4) of the Gaia mission, expected in late 2026, will provide exquisite parallaxes of Milky Way (MW) Cepheids, enabling the calibration of their PL relation with sub-percent accuracy. As with all stars in the MW, the light from Cepheids is affected by the interstellar medium (ISM), making specific corrections necessary.
As parallax uncertainties decrease, extinction becomes one of the dominant sources of systematic error in the calibration of the PL relation. Furthermore, errors in Cepheid extinction introduce scatter and systematics in the calibration of their surface brightness-color relations \citep[SBCRs,][]{Kervella2004_SBCR, Storm2011, Bailleul2025, Bailleul2026} used to derive the angular diameters from their photometry. Cepheid SBCRs are used in the parallax-of-pulsation (PoP) method, also known as the Baade-Wesselink method \citep{Baade1926, Wesselink1969}, to determine their distances, which can also be used to calibrate the PL relation.
The effect of extinction corrections is considerable due to the strong spatial variability of dust across the MW and the intrinsic variability of Cepheids, which complicates direct reddening estimates due to changing luminosity and temperature.

Starlight is scattered and absorbed by interstellar dust mostly in the ultraviolet, optical, and near-infrared domains, with the extinction being stronger at shorter wavelengths, inducing the reddening effect.
The total extinction $A_{\lambda} = m_{\lambda, \mathrm{obs}} - m_{\lambda, 0}$ gives the wavelength-dependent attenuation of the observed brightness $m_{\lambda, \mathrm{obs}}$ compared to the intrinsic magnitude $m_{\lambda, 0}$. Traditionally, extinction $A_{\mathrm{V}}$ is expressed in the visible band. The wavelength dependence of $A_{\lambda}$ is determined by the dust properties, which vary widely within the MW, and of which we generally do not have exact knowledge. 
The heterogeneous distribution of dust properties is reflected in dust extinction curves, which show variation in their wavelength-dependent shapes across the MW. However, most of them are strongly correlated with one parameter, the total-to-selective extinction ratio $R_{\mathrm{V}} = A_{\mathrm{V}}/E_{\mathrm{B-V}}$, where $E_{\mathrm{B-V}}$ is the color excess $E_{\mathrm{B-V}} = (m_{\mathrm{B}} - m_{\mathrm{V}}) - (m_{\mathrm{B,0}} - m_{\mathrm{V,0}}) = A_{\mathrm{B}} - A_{\mathrm{V}}$. The $R_{\mathrm{V}}$ is commonly viewed as a proxy for the average grain size along a sightline, with the larger $R_{\mathrm{V}}$ indicating larger grains \citep{Zelko2020}. The first MW extinction relation depending on $R_{\mathrm{V}}$ was calibrated by \citet{Cardelli1989}, and later was followed by many more \citep[e.g.,][]{ODonnel1994, Fitzpatrick1999, Valencic2004, Fitzpatrick2007, Fitzpatrick2019, Wang2019, Decleir2022} with diverse wavelength coverages, and large samples of both spectra and photometry used to construct the relations.

\citet{Gordon2023} derived an $R_{\mathrm{V}}$-dependent extinction relation by combining spectra from recent studies and fitting extinction simultaneously over an extensive range from $912~\AA$ to $32~\mu$m. They resolve the continuum and the features from the ultraviolet (UV) to infrared (IR) of the extinction curves as a function of $R_{\mathrm{V}}$. Although $R_{\mathrm{V}}$ is an efficient parameter for counting the variable behavior of the extinction curve within the MW, the relations based on $R_{\mathrm{V}}$ only describe the average behavior of the dust extinction.

The average $R_{\mathrm{V}}$ for the diffuse medium in the MW is around $3.1$ \citep[e.g.,][]{Cardelli1989, Fitzpatrick1999, Wang2019, Zhang2025}. It is widely used in modeling Galactic extinction, although $R_{\mathrm{V}}$ varies significantly depending on the local environment. \citet{Zhang2025} derived three-dimensional maps of $R_{\mathrm{V}}$ by measuring the extinction curve of 130 million stars from low-resolution spectra. Within $4\,$kpc of the Sun, they found a median $R_{\mathrm{V}}=3.11$ with a $1\sigma$ distribution ranging from 2.60 to 3.88. To account for this in this work, we will consider the variation in $R_{\mathrm{V}}$ in addition to $A_{\mathrm{V}}$ in the dust extinction modeling.

It should be noted that the extinction measured in physical $B$ and $V$ bandpasses depends on the source spectrum and the filter transmission functions. Extinction relations are generally defined in terms of the monochromatic wavelength dependence of extinction, and therefore do not strictly reproduce band-integrated extinctions for arbitrary sources. Subsequently, the parameters $A_{\mathrm{V}}$ and $R_{\mathrm{V}}$ in these relations do not exactly correspond to the observed broadband extinctions. They rather parameterize the normalization and shape of the underlying extinction curve.

A conventional way to estimate stellar extinction is to compare the observed spectral energy distribution (SED) with the intrinsic SED based on empirical templates or stellar atmosphere models.
However, changing luminosity makes it difficult to define a single magnitude value for a Cepheid and, therefore, to assemble a stable SED. Reddening values for Cepheids are traditionally derived using phase-averaged magnitudes and colors from light curves \citep[e.g.,][]{Fernie1995, Laney2007, Monson2011}, but variability inevitably introduces uncertainty. Furthermore, most Cepheids are binaries, often with a hot main-sequence companion \citep{Kervella2019, Evans2022, Karczmarek2022, Gallenne2013a, Gallenne2014, Gallenne2015A} that potentially introduces a bias in the B-V color of the Cepheid.

The observed extinction of Cepheids can be mitigated using the Wesenheit indices \citep{Madore1982}. They are a combination of magnitudes, for example, $W_{\mathrm{VB}}=V-R(B-V)$, where $V$ and $B$ are the apparent magnitudes of the Cepheid and $R=A_{\mathrm{V}}/(A_{\mathrm{B}}-A_{\mathrm{V}})$. The Wesenheit indices are particularly advantageous for Cepheids in distant galaxies, where individual stellar extinction cannot be determined. The same method can be consistently applied to both Galactic and extragalactic Cepheids, which is why the Wesenheit index is used for measuring the Hubble constant
from the cosmic distance ladder \citep{Riess2022, Riess2024}. Although $W$ essentially eliminates the effects of extinction on the PL relationship, its definition based on broadband photometry still depends on the effective temperature of the source ($T_{\mathrm{eff}}$), which introduces an uncertainty for Cepheids with changing $T_{\mathrm{eff}}$. Recently, \citet{Skowron2026} and \citet{Wang2026} showed that the Wesenheit indices depend significantly on the adopted value of $R_{\mathrm{V}}$, and assuming a constant $R_{\mathrm{V}}$ for the MW Cepheids introduces systematic error in the resulting Cepheid distances, particularly in the optical regime. These uncertainties may become a limiting factor when calibrating the PL relation with Gaia parallaxes.

An alternative approach to estimate Cepheid extinction is to retrieve it at the Cepheid's location from three-dimensional dust maps. These maps aim to reconstruct the 3D distribution of interstellar extinction in the Galactic neighborhood as a function of sky position and distance. Recent efforts, e.g., Bayestar19 \citep{Green2014, Green2019}, G-Tomo \citep{Lallement2022, Vergely2022}, and DECaPS 3D dust map \citep{Zucker2025}, use hundreds of millions of stars, often with parallaxes from the Gaia catalogs \citep{Gaia2016} and photometry from large surveys such as Pan-STARRS 1 \citep{Chambers2016}, 2MASS \citep{Skrutskie2006}, and VVV \citep{Minniti2010}.
While they are remarkable tools for extinction corrections in various fields and for the study of the Local Arm structure, the 3D dust maps are limited in sky coverage, depth, accuracy, and resolution due to data availability and the fact that they derive stellar class and reddening solely from photometry.
Therefore, they are insufficient for obtaining precise extinction values to get the full potential of the Gaia parallaxes when recalibrating the PL relation. Nevertheless, dust maps remain an important source for comparison.

In this work, we introduce an original strategy to determine the extinction of 16 MW Cepheids from the extinction of the surrounding field stars. These field stars are carefully selected from the Gaia DR3 catalog based on their coordinates and parallax relative to the Cepheid. We determine the field star and Cepheid extinction following these steps:
\begin{enumerate}[itemsep=8pt]
    \item We observe the field stars with the VLT/FLAMES spectrograph and derive their stellar atmospheric parameters ($T_{\mathrm{eff}}, \log g, \left[\mathrm{M}/\mathrm{H}\right]$) from the spectra.
    \item We build the field star SEDs from Gaia DR3 XP spectra and photometry from the 2MASS and AllWISE surveys and estimate their individual extinction by fitting atmosphere models with stellar parameters obtained in step 1.
    \item We determine Cepheid $A_{\mathrm{V}}$ and $R_{\mathrm{V}}$ by interpolating the extinction values in three dimensions.
\end{enumerate}
By estimating $A_{\mathrm{V}}$ from the observations of field stars, we avoid the complications related to the Cepheid pulsation. Moreover, we break the usual degeneracy between the effective temperature and extinction in SED fitting by adopting stellar parameters from spectroscopy. With this method, we achieve extinction estimates for Galactic Cepheids with high accuracy. The purpose of this paper is to introduce the method with a pilot sample of 16 MW Cepheids, covering a wide range of distances, pulsation periods, and extinction. In upcoming publications, we will complement this sample with new observations of 84 Cepheid fields. We will use the resulting extinction estimates of these 100 Cepheids combined with Gaia DR4 parallaxes to calibrate the PL relation with improved precision.

\section{Observations and data query}
We observed the field stars around 16 classical Milky Way Cepheids with the multi-object spectrograph VLT/FLAMES. Astrometric and photometric data were gathered from three large surveys with a good coverage of our sample:
\begin{itemize}
    \item Gaia Data Release 3 \citep[Gaia DR3,][]{Gaia2023}
    \item Two Micron All Sky Survey \citep[2MASS,][]{Skrutskie2006}
    \item AllWISE Data Release of the Wide-field Infrared Survey Explorer (WISE) observations \citep{Cutri2013}
\end{itemize}

\subsection{Cepheid and field star samples} \label{sec:star_selection}
Our pilot sample includes 16 Galactic Cepheids pulsating in the fundamental mode (Table \ref{tab:cepheids}). The sample is part of a larger program of 100 Cepheids selected from the compilation by \citet{Groenewegen2018} with 450 well-established classical Cepheids based on their suitability as PL relation calibrators. In the selection, we considered their presence in previous catalogs and PL-relation studies \citep{Genovali2014, Ripepi2019, Groenewegen2018, Breuval2021, Trahin2021, Riess2021, Narloch2023}, as well as the quality of their distance measurements (i.e., parallax with a sufficient signal-to-noise ratio in Gaia DR3). In addition, to validate our methodology under diverse conditions, we ensured a variety of distances, pulsation periods, and literature extinction values. The 16 Cepheids are representative of the full sample with $d=340-4700\,$pc ($\varpi=0.19-2.93\,$mas), $P=5-41\,$days and $E_{\mathrm{B-V}}=0.06-0.87$, and were therefore chosen for our FLAMES pilot program.

Our approach is based on identifying the stars closest to the Cepheids, which we will refer to as field stars. Their selection from the Gaia DR3 catalog was based on both the angular and radial distances to the Cepheids, thus accounting for their separation in three-dimensional space. The angular separation was limited to 12.5 arcminutes due to the field of view of the VLT/FLAMES instrument. 
We typically limited the radial distance to $\pm20-30\%$ of the Cepheid parallax,
but also included stars farther away to maximize the number of optical fibers used in the FLAMES observations. The most flexibility in the line-of-sight distance was given to the closest ($\beta$ Dor, $\ell$ Car) and most distant (VZ Pup, LS Pup) targets since the field stars are strongly concentrated on one side of the Cepheid.
The magnitude was limited by the instrument sensitivity to $G\leq17\,$mag. Additionally, we used some of the catalog's quality flags, requiring the parallax-over-error ratio (RPlx) to be above 10 and the renormalized unit weight error (RUWE) below 1.4. The RUWE parameter is used to assess the quality of the astrometric solution, and a value greater than 1.4 could indicate binarity or calibration problems \citep{Lindegren2021}. We adopted Gaia DR3 parallaxes as distance estimates for all 16 Cepheids, even when the RUWE number was greater than 1.4. Although the likely Cepheid multiplicity could bias the parallax based on a single-star astrometric solution, Gaia DR3 provides the most precise and recent distance estimates for most Galactic Cepheids.

\begin{figure}
\centering
\includegraphics[width=8.8cm]{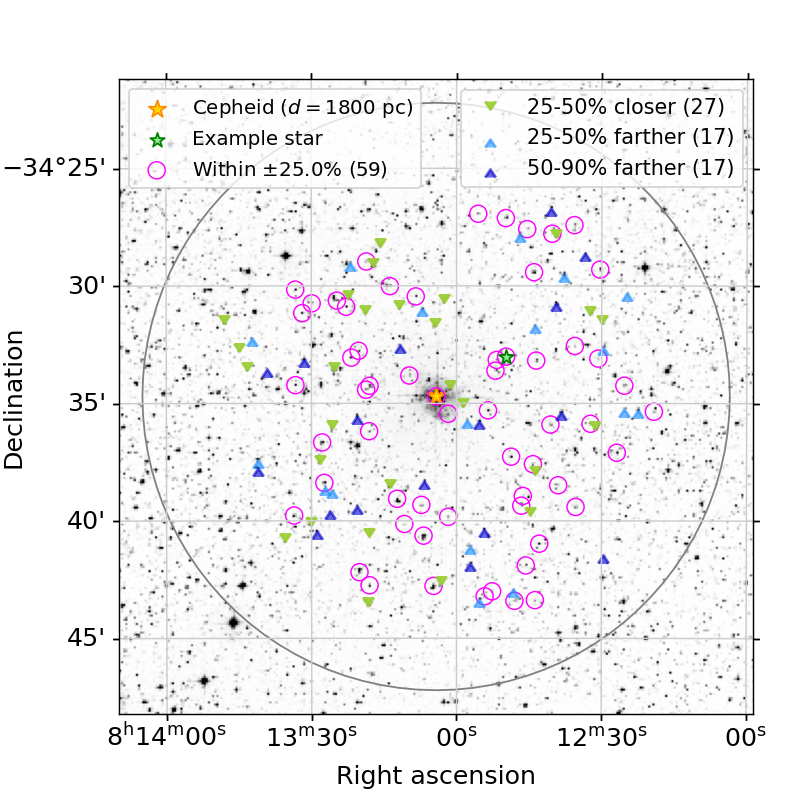}
   \caption{Selected field stars around the Cepheid RS Pup. The stars within $\pm25\%$ of the RS Pup distance are marked with fuchsia circles, and the closer and more distant ones with green and blue triangles. The big gray circle shows the field of view of the FLAMES spectrograph. The field star marked with a green star (Gaia DR3 5546480122794409472) is used as an example in the spectral and SED fitting (Fig. \ref{fig: giraffe_spectrum} and \ref{fig: SED}).}
      \label{fig: RS Pup chart}
\end{figure}

A map of field stars for the Cepheid RS Pup is shown in Fig. \ref{fig: RS Pup chart} with the green star indicating the position of Gaia DR3 5546480122794409472, which we will later use as an example star to demonstrate the different steps in the analysis. We queried the Gaia DR3 catalog \citep[I/355,][]{Gaia_Vizier_2022} through the VizieR web service and accessed it through the Python interface astroquery \citep{Ginsburg2019}. On average, we selected 120 field stars per Cepheid, adding up to a total of 1918 stars spread across the 16 fields with a mean $G$ magnitude of $14.5$.

\subsection{VLT/FLAMES observations}
The spectra of the Cepheid field stars were observed with the multi-element FLAMES spectrograph \citep{Pasquini2002} installed at the Nasmyth A focus of the Very Large Telescope's (VLT) second $8.2$-m unit telescope (UT2). FLAMES feeds two different spectrographs: the medium-high-resolution (R = 5500-65000) GIRAFFE instrument and the red arm of the high resolution (R $\sim$ 47000) UVES spectrograph \citep{Dekker2000}.
We used a configuration combining both spectrographs, with an average of 115 field stars observed with GIRAFFE and 3-6 stars with UVES. The GIRAFFE spectra were obtained in high-resolution mode with three settings 
\begin{itemize}
    \item HR15N ($647-679\,$nm, R $\sim$ 19200), 
    \item HR18 ($747-789\,$nm, R $\sim$ 20150),
    \item HR21 ($848-900\,$nm, R $\sim$ 18000). 
\end{itemize}
The UVES red arm provided a wide spectral range from $480\,$nm to $1100\,$nm, observed in two setups with central wavelengths of $580$ and $860\,$nm.
Each Cepheid field was observed three times with the three GIRAFFE settings and an exposure time of approximately $360\,$s. With UVES, we had six exposures of $360\,$s with the $580\,$nm setup and three with the $860\,$nm setup. 
With these observations, we reached a signal-to-noise ratio (S/N) of about 40 for the GIRAFFE spectrum of a typical field star with $G\sim14.5$. For UVES, we estimated the flux uncertainties in an alternative way that provides only conservative estimates of their S/N (see Sec. \ref{sec: data reduction} for details).

\subsection{Data reduction} \label{sec: data reduction}
We reduced the observations with the ESO GIRAFFE and UVES-FIBER data-reduction pipelines \citep{Ballester2000} in the EsoReflex environment \citep{Freudling2013}. We cleaned the GIRAFFE spectra from cosmic rays (CRs) using the Python package Astro-SCRAPPY \citep{McCully2019}, which is based on the L.A.Cosmic algorithm \citep{vanDokkum2001} and is integrated into EsoReflex. For UVES, CR cleaning was included in the pipeline.

The UVES flux uncertainties provided by the pipeline are unrealistically low for our observations, resulting in an S/N of the order of $10^3-10^4$. Instead of adopting the pipeline uncertainties, we estimated the error bars as the root mean square (RMS) of the normalized flux in the continuum regions. This resulted in conservative uncertainties for the UVES spectra with an S/N below 10 when the expected value would be around $30$ for a typical field star with $G\sim14.5$, with most of the chosen UVES stars brighter than that.

\subsection{Data query} \label{sec: data query}
We included archival spectrophotometric and photometric data from Gaia DR3 \citep{Gaia2023}, 2MASS \citep{Skrutskie2006}, and AllWISE \citep{Cutri2013}.We selected these catalogs because of their coverage of our sample and the quality of the photometric data in terms of absolute calibration.

\subsubsection{Gaia DR3}
The third data release \citep[DR3][]{Gaia2023} from the European Space Agency's Gaia mission \citep{Gaia2016} includes astrometry and photometry for more than 1.4 billion sources. In our work, we leverage the astrometry (position and parallax) to localize the Cepheid field stars (see Sec. \ref{sec:star_selection}). Gaia provides exceptional precision for distances, with a median parallax uncertainty of $0.02\,$mas for the field stars in our sample.

We also used the Gaia observations in the construction of the field star SEDs (see Sec. \ref{sec: SEDs}). The Gaia photometric bands G ($330-1050\,$nm), BP ($330-680\,$nm), and RP ($630-1050\,$nm) are broad, making them suboptimal for SEDs.
Hence, we opted for the low-resolution BP/RP spectra (also referred to as XP spectra) over the Gaia photometry. The XP spectra were first published in DR3, where the mean spectra are generated from multiple observations (transits) of the same object. They are provided for about 220 million sources in DR3 and are available for all stars with G magnitude brighter than 17.65 \citep{DeAngeli2023}. Although our entire sample is brighter than this limit, XP spectra were not available for a few field stars around the following Cepheids: T~Mon (17 stars), CV Mon (12 stars), and LS Pup (1 star). For these 30 stars, we used Gaia (G, BP, RP) photometry instead.

XP spectra cover the range $336-1020\,$nm with a resolving power of $R\sim20-70$. The resolution is too low to derive accurate atmospheric parameters from spectral lines, but it is sufficient for the general shape of the SED. The optical range is sensitive to extinction effects, which makes the XP spectra ideal for constraining both the amount of extinction and the shape of the extinction curve. However, we include the XP spectra only between $400-1000\,$nm due to the known systematic error and the overall higher uncertainty of the observations outside this range \citep{Montegriffo2023, Babusiaux2023}.

\subsubsection{2MASS}
The Two Micron All Sky survey (2MASS) imaged the NIR sky between 1997 and 2001 with two 1.3-m telescopes \citep{Skrutskie2006}. The three bands $J$ ($1.25~\mu$m), $H$ ($1.65~\mu$m), and $K_{\mathrm{s}}$ ($2.17~\mu$m) reach sensitivities of 15.8, 15.1, and 14.3 mag with a $10\sigma$ point-source detection level. We included 2MASS in the stellar SEDs to anchor them in the NIR, where the impact of extinction is lower than in the optical range.

We queried for the 2MASS photometry by cross-identifying our field stars with the 2MASS VizieR catalog (II/246) based on their Gaia DR3 coordinates\footnote{We used the astroquery.xmatch tool for the cross-identification between the catalogs.}.
The $JHK_{\mathrm{s}}$ photometry is available for $97.3\%$ of the sample. For the $K_{\mathrm{s}}$ band that is most affected by the thermal background, the mean uncertainty within our sample is $0.04\,$mag. We checked photometric quality flags and included only clear detections with S/N$~>7$. For $4.7\%$ of the 1867 stars, at least one measurement did not meet this criterion and was removed.

\subsubsection{AllWISE}
The Wide-field Infrared Survey Explorer \citep[WISE][]{Wright2010} was NASA's survey mission active from 2009 to 2011. WISE mapped the infrared sky in four bands $W1$ (3.4$~\mu$m), $W2$ (4.6$~\mu$m), $W3$ (12$~\mu$m) and $W4$ (22$~\mu$m). The AllWISE source catalog combines data from cryogenic and post-cryogenic survey phases with improved sensitivity and accuracy compared to earlier data releases \citep{Cutri2013}.

We cross-identified the field stars in the AllWISE catalog (II/328) and found matches for $83.3\%$. Due to the high uncertainty associated with the $W3$ and $W4$ magnitudes, we included only the $W1$ and $W2$ bands in the stellar SEDs when available. In these mid-infrared (MIR) wavelengths, the reddening effect is minor, which helps identify the unreddened stellar continuum. We excluded at least one measurement for $3.2\%$ of the 1598 stars due to quality flags that indicated strong contamination or S/N$~<10$.

\section{Methods}

\subsection{Atmospheric parameters from spectra} \label{sec: param from spectra}

We estimated the stellar atmospheric parameters, i.e., effective temperature ($T_{\mathrm{eff}}$), surface gravity ($\log g$), and metal abundance (often referred to as metallicity, $\left[\mathrm{M}/\mathrm{H}\right]$) of the field stars by fitting the VLT/FLAMES observations with synthetic stellar spectra. For this task, we used the iSpec spectral tool \citep{Blanco-Cuaresma2014b, Blanco-Cuaresma2019}. We also fitted alpha enhancement ([$\alpha$/Fe]) and projected rotational velocity ($\upsilon \sin{i}$), and estimated values for microturbulent ($V_{\mathrm{mic}}$) and macroturbulent ($V_{\mathrm{mac}}$) velocities. The limb-darkening coefficient was fixed to 0.6 for the spectral fitting, as it acts as a simple scaling factor and has no impact on the relative depths of the spectral features. Resolving power $R$ was set to the estimated value for the specific instrument setup, i.e., $R=18000$ for GIRAFFE and $R=47000$ for UVES. The spectral analysis routine described below was customized for our observations to find reliable parameter values through trial and error.

To fully benefit from the observations, we aimed to use the entire wavelength range of the observed spectra. However, the GIRAFFE HR18 spectra were greatly contaminated by telluric absorption of atmospheric $\mathrm{O}_2$ as well as the UVES spectra with the setup centered at $860\,$nm by both $\mathrm{H}_2\mathrm{O}$ and $\mathrm{O}_2$ \citep[e.g., Fig.~1 in][]{Smette2015}. We attempted to correct for this telluric absorption using the Molecfit software tool \citep{Smette2015, Kausch2015} but found the quality of the corrected spectra to be still insufficient for our purposes. Even so, the remaining HR15N and HR21 spectra (excluding $891-900\,$nm due to $\mathrm{H}_2\mathrm{O}$ lines) for GIRAFFE and the UVES 580-nm setup spectra were sufficient to obtain reliable atmospheric parameter values.

\begin{figure}
\centering
    \includegraphics[width=8.8cm]{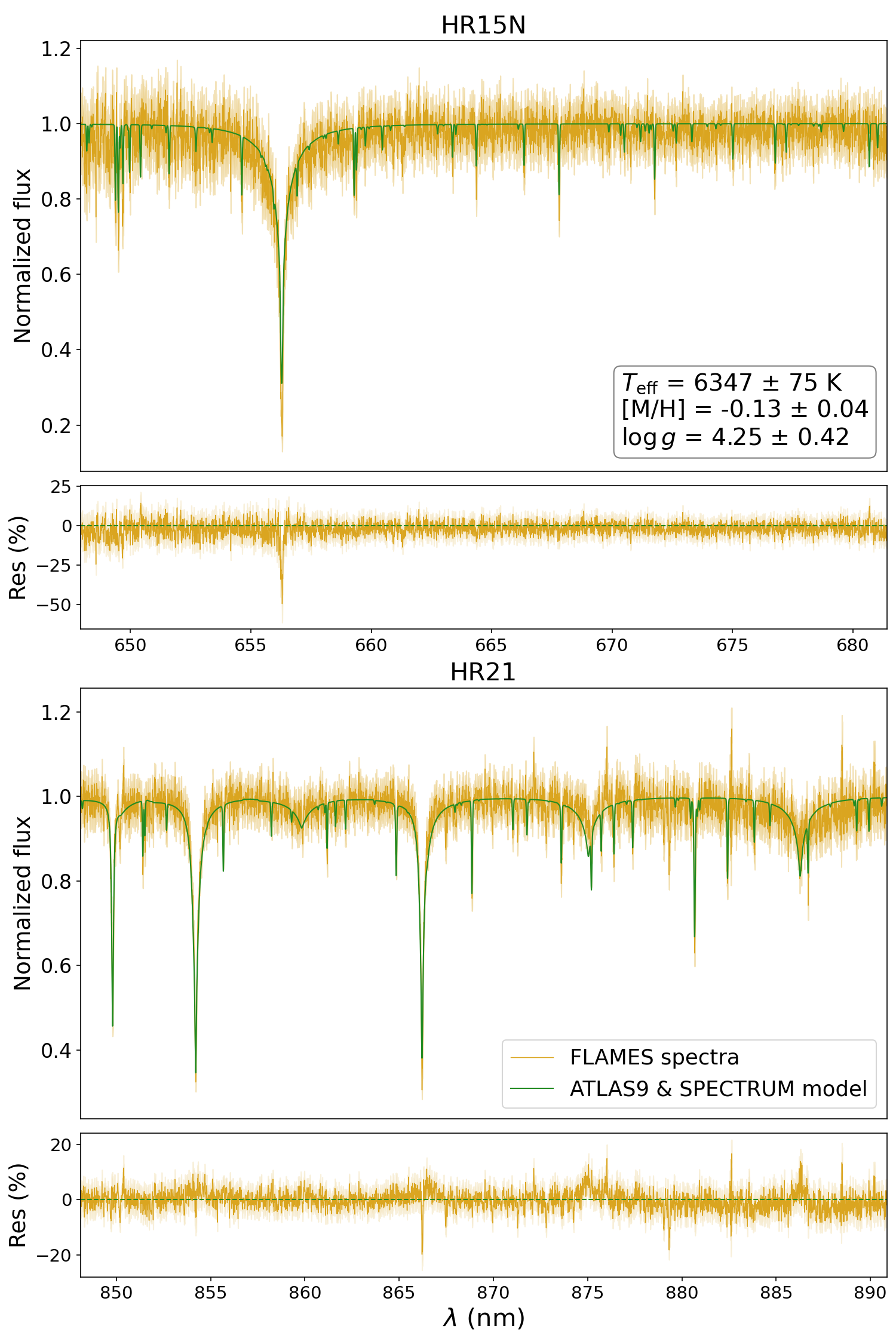}
   \caption{Spectral fit of Gaia DR3 5546480122794409472, a field star of Cepheid RS Pup. The spectra were observed with the FLAMES/GIRAFFE spectrograph (yellow) and fitted with model spectra (green) based on ATLAS9 atmosphere models built with the radiative transfer code \texttt{SPECTRUM}.
           }
      \label{fig: giraffe_spectrum}
\end{figure}

With iSpec, we co-added the exposures to increase the signal-to-noise ratio. We normalized the spectra by applying median and maximum filters with window steps of 0.2 and 1.0 nm, respectively, and fitting a second-degree B-spline to find the continuum while ignoring strong lines. The spectrum was normalized by dividing the observation with the fitted continuum. 
We merged the different wavelength regions to fit the parameters simultaneously. Radial velocities were found and corrected by cross-correlating the normalized spectra with a template \cite[NARVAL solar spectrum, see][]{Blanco-Cuaresma2019}. 

The atmospheric parameters were determined by fitting the observations with synthetic stellar spectra. These spectra were calculated with the radiative transfer code \texttt{SPECTRUM} \citep{Gray1994} by interpolating the ATLAS9 stellar atmosphere model grid \citep{Castelli2003}. The fitting to the observations was based on $\chi^2$ minimization with the non-linear least squares fitting program \texttt{MPFIT} \citep{Markwardt2009}, and the parameter errors were derived from the covariance matrix. The atomic absorption lines that were used to find the parameters were taken from the extensive Gaia-ESO Survey (GES) line list \citep[version 6,][]{Heiter2021}. This list is based on UVES observations, hence we adopted it for our UVES spectra as-is. For GIRAFFE observations with lower resolution, we cross-matched the GES v6 list with our best-quality (i.e., high S/N) GIRAFFE spectra to create a custom-made list.
The observed and fitted GIRAFFE spectra of the example star with $G=15.4$, which represents the dimmer end of our field star sample, are shown in Fig. \ref{fig: giraffe_spectrum}. An example of the UVES spectrum is given in Appendix \ref{appendix: UVES spec}.

The UVES spectra include strong absorption features that help break the degeneracies between the parameters, namely the broad wings of H$\alpha$ ($652-660\,$nm) and H$\beta$ ($483-490\,$nm), the Mg I b triplet ($516-519\,$nm), and the numerous Fe I/II lines. The GIRAFFE spectra also feature H$\alpha$, which helps constrain $T_{\mathrm{eff}}$, and Fe lines for $\left[\mathrm{M}/\mathrm{H}\right]$. However, the GIRAFFE spectral range does not include the $\log g$-sensitive Mg I b triplet, which is why iSpec had difficulty constraining $\log g$. In most cases, $\log g$ was likely overestimated, usually above $4.5$, often pushed to its maximum of 5.0 without uncertainty estimates.

Due to the difficulty in constraining $\log g$, we opted to estimate it separately, using the theoretical \texttt{MESA} Isochrone and Stellar Tracks (MIST)
models \citep{Dotter2016, Choi2016, Paxton2011, Paxton2013, Paxton2015} computed for solar metallicity $[Fe/H]=0.0$ and zero rotation $\upsilon/\upsilon_{\mathrm{crit}}=0$. The MIST isochrones relate, e.g., stellar surface parameters and Gaia photometry, which we used to compute Gaia color-magnitude coordinates ($BP-RP$, $G$), and to find the corresponding theoretical $\log g$. We trained a k-nearest-neighbor (k-NN) regressor\footnote{We used the Python function sklearn.neighbors.KNeighborsRegressor.} with these values, choosing $k=5$ and distance weighting. To make the observed Gaia magnitudes of the field stars compatible with the theoretical isochrones, we corrected them for extinction using the values from the 3D extinction maps Bayestar \citep{Green2019} and DECaPS \citep{Zucker2025}, and the \citet{Gordon2023} extinction relation (see Sec. \ref{sec: SEDs} for more about modeling the extinction). We then estimated the absolute magnitudes based on the Gaia parallaxes and used them to predict the field star $\log g$ from the isochrones.
We fixed $\log g$ to this value when fitting synthetic spectra.
We set the uncertainties conservatively to $\pm10\%$ of the $\log g$ value. See Appendix \ref{appendix: logg} for a comparison of $\log g$ estimated from the MIST isochrones and the UVES spectra.

We fitted the parameters $T_{\mathrm{eff}}$ and $\left[\mathrm{M}/\mathrm{H}\right]$, and their initial values, when available, were taken from the Gaia DR3 catalog, which includes stellar parameters for 471 million sources estimated with the General Stellar Parameterizer from Photometry \citep[GSP-Phot,][]{Andrae2023}. The GSP-Phot is based on the simultaneous fitting of the low-resolution XP spectrum, parallax, and apparent G magnitude with a Bayesian forward-modeling approach. The results from this method depend strongly on the data quality, particularly on the parallax measurements, which is why we do not use the GSP-Phot stellar parameters directly for our field stars. However, as initial values, they help our spectral fits converge faster and are available for $85\%$ of our total field star sample. For the remaining $15\%$, we chose $T_{\mathrm{eff}}=5500\,$K and $\left[\mathrm{M}/\mathrm{H}\right]=0.0$ as initial values.

Velocities $V_{\mathrm{mic}}$ and $V_{\mathrm{mac}}$ were estimated and fixed from empirical relations considering $T_{\mathrm{eff}}$, $\log g$ and $\left[\mathrm{M}/\mathrm{H}\right]$. The relations were constructed based on the UVES Gaia-ESO survey \citep{Gilmore2012, Randich2013}, the Gaia FGK benchmark stars \citep{Jofre2014}, and globular cluster data from the literature, and are implemented within iSpec \citep{Blanco-Cuaresma2014b, Blanco-Cuaresma2014a}. The parameter $\upsilon \sin{i}$ was kept free in the fit. With the macroturbulence velocity fixed to the value obtained from the relation, and taking into account the resolving power of the instruments, $\upsilon \sin{i}$ is estimated from the broadening of the spectral lines without strong degeneracies. We added a systematic error of $1\%$ of each parameter value to the statistical error bars estimated by iSpec to account for, e.g., uncertainties associated with the continuum normalization. The values and uncertainties found for all parameters are given in the supplementary material.

The fitting process was automatized, and we flagged fits with high $\chi^2$ values or missing parameter uncertainty estimates. We visually inspected the flagged fits and discarded the stars for which it was evidently not possible to determine accurate parameters. These included stars with low S/N spectra, binaries with double spectral features, fast-rotating stars with strong line blending, and, in a few cases, stars with problems in their observations. For instance, we found an exceptionally low metal abundance ($\left[\mathrm{M}/\mathrm{H}\right]<-3$) with high projected rotational velocity ($\upsilon \sin{i} > 100~\mathrm{kms}^{-1}$) for 19 stars, suggesting that the metal lines had blended together. Since we suspect the low $\left[\mathrm{M}/\mathrm{H}\right]$ not to be physical, we excluded these stars.
Additionally, we checked the Gaia DR3 variability flag \citep{Rimoldini2023} and excluded stars classified as RS Canum Venaticorum type variables (RS, 7 targets) and eclipsing binaries (ECL, 11 targets).
About half of the field stars were excluded for $\beta$ Dor as many of them were faint ($G>16$) and had low S/N, so that out of the 112 observed stars we could derive the parameters for 60. In total we could determine the stellar parameters for $1609$ field stars, leaving us on average $89\%$ of the observed stars per Cepheid field.

\subsection{Field star extinction from SEDs} \label{sec: SEDs}

We built the SEDs of the field stars from large spectroscopic and photometric surveys. We fitted the SED of each field star using ATLAS9 atmosphere models and parameterized extinction curves to estimate the individual extinction of the field stars.
The stellar atmospheric parameters determined in the spectral fitting were incorporated as Gaussian priors in the SED fits.

The low-resolution Gaia XP spectra set the basis, covering the optical range between $400$ and $1000\,$nm. We smoothed the spectra to better match the model resolution by applying a Gaussian kernel\footnote{We used the Python function scipy.ndimage.gaussian\_filter1d.} with a width of $4\,$nm. This was complemented by 2MASS ($JHK$) and AllWISE ($W1$ and $W2$) photometry in the NIR and MIR.
No XP spectrum, 2MASS or AllWISE photometry was found for one and two stars in the fields of T~Mon and CV~Mon, respectively, hence they were excluded from the analysis.
For the photometric bands, we retrieved the filter zero points and transmission curves from the Spanish Virtual Observatory (SVO) Filter Profile Service \citep{Rodrigo2012, Rodrigo2024, Rodrigo2020}. We converted the magnitudes with the error estimates to flux densities using the respective zero points.

\begin{figure} 
\centering
\includegraphics[width=8.8cm]{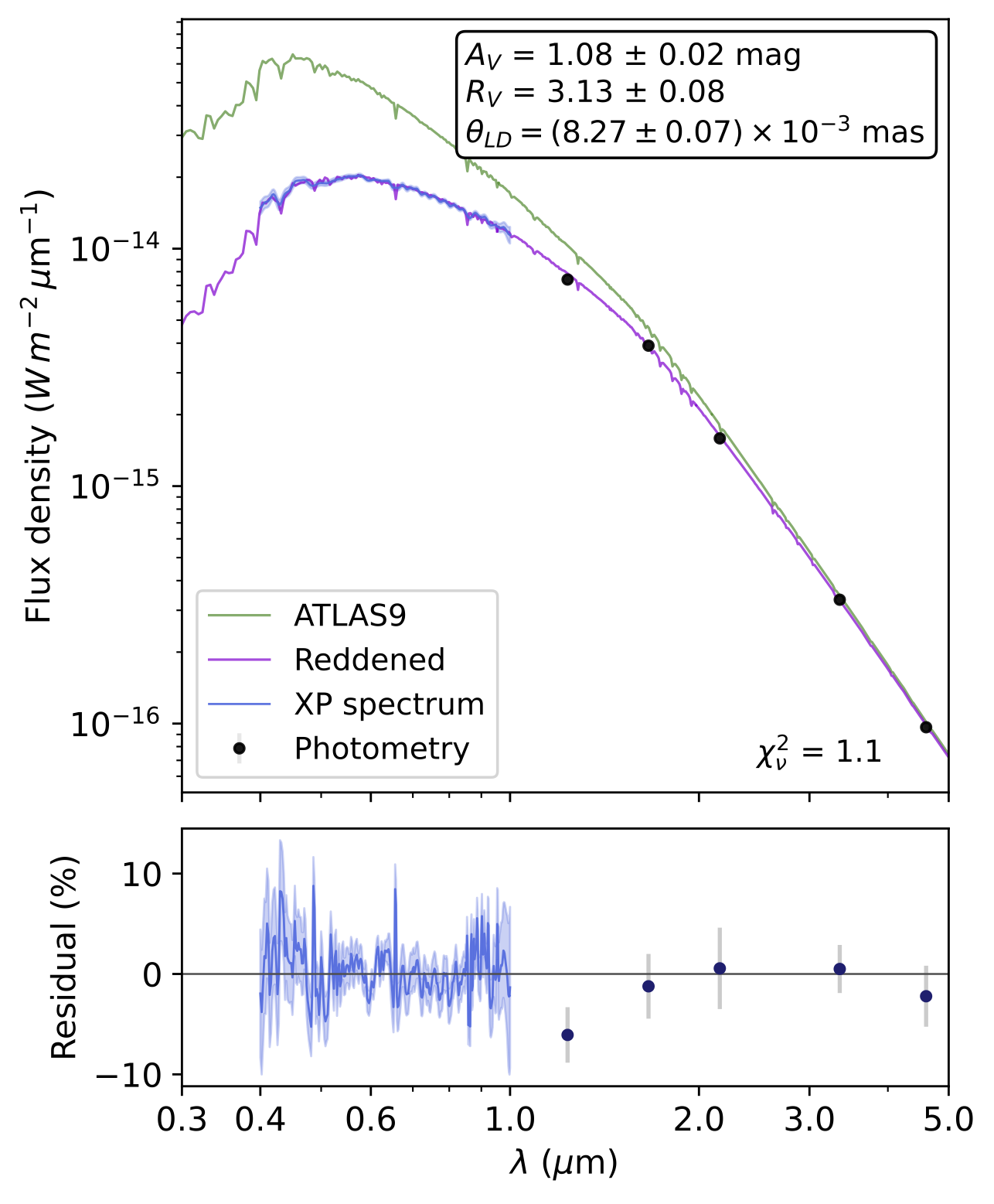}
   \caption{SED of Gaia DR3 5546480122794409472, a field star of Cepheid RS Pup (same as in Fig \ref{fig: giraffe_spectrum}), fitted with ATLAS9 atmosphere models and G23 extinction relation. The parameter values resulting from the fit are on the top right, while priors for $T_{\mathrm{eff}}$, $\left[\mathrm{M}/\mathrm{H}\right]$, and $\log g$ were introduced from the spectra or isochrones.
           }
      \label{fig: SED}
\end{figure}

The synthetic SED was constructed from the ATLAS9 atmospheres \citep{Castelli2003} by interpolating the model grid based on the given set of parameters. The free parameters were $A_{\mathrm{V}}$, $R_{\mathrm{V}}$, and the limb-darkened angular diameter $\theta_{\mathrm{LD}}$, while we introduced the values of $T_{\mathrm{eff}}$, $\log g$ (UVES stars), and $\left[\mathrm{M}/\mathrm{H}\right]$ from the spectral fitting, and $\log g$ (GIRAFFE stars) from the MIST isochrones, as priors.
We performed least-squares fitting using the implementation in the SciPy library \citep{Virtanen2020}.
To prevent the densely sampled XP spectra from dominating the fit, we applied dataset-level weighting by normalizing the spectroscopic and photometric contributions by their respective numbers of data points. Therefore, the weighted $\chi^2$ statistic constrained by the observations is 
\begin{align}
    \chi_{\mathrm{obs}}^2 = \frac{1}{N_{\mathrm{spec}}}\sum_i^{N_{\mathrm{spec}}}\left(\frac{F^{\mathrm{obs}}_{\mathrm{spec,}i} - F^{\mathrm{mod}}_{\mathrm{spec,}i}}{\sigma_{\mathrm{spec,}i}}\right)^2 + \\ \notag
    \frac{1}{N_{\mathrm{phot}}}\sum_j^{N_{\mathrm{phot}}}\left(\frac{F^{\mathrm{obs}}_{\mathrm{phot,}j} - F^{\mathrm{mod}}_{\mathrm{phot,}j}}{\sigma_{\mathrm{phot,}j}}\right)^2,
\end{align}
where $N_{\mathrm{spec}}$ and $N_{\mathrm{phot}}$ are the numbers of spectroscopic and photometric data points, respectively, $F^{\mathrm{obs}}$ and $\sigma_{\mathrm{obs}}$ are the observed flux density and its uncertainty, and $F^{\mathrm{mod}}$ is the modeled flux density.
To this we added Gaussian penalties so that the minimized quantity is
\begin{align}
    \chi_{\mathrm{fit}}^2=\chi_{\mathrm{obs}}^2+\sum_i\left(\frac{p_i-p_{i, 0}}{\sigma_{p,i}}\right)^2,
\end{align}
where $p_{i, 0}$ is the value and $\sigma_{p,i}$ the uncertainty of each parameter derived from the spectra or isochrones. 
To evaluate the effect of reddening, we reddened the synthetic SED with the \citet{Gordon2023} extinction relation (G23) parameterized by $A_{\mathrm{V}}$ and $R_{\mathrm{V}}\in[2.3,5.6]$. The extinction curves were accessed with the Python package dust\_extinction \citep{Gordon2024} and applied to the SED with the help of synphot.reddening module \citep{synphot2018}.
The fitted SED of the example star is given in Fig. \ref{fig: SED} with atmospheric parameters taken from the spectral fit in Fig. \ref{fig: giraffe_spectrum}.

To optimize the fitting process, we set initial values for the free parameters.
We took the initial extinction $A_{\mathrm{V}}$ from the 3D dust extinction maps Bayestar19 of the northern sky \citep{Green2014, Green2019} and DECaPS of the southern Galactic plane \citep{Zucker2025} based on the field star's position in space, as listed in the Gaia DR3 catalo (RA, DEC, Plx). The query was performed with the Python package dustmaps \citep{Green2018}, and we retrieved only the mean values for the field stars from the probabilistic maps. The extinction in the maps is given as color excess $E_{\mathrm{B-V}}$, and we converted the values to $A_{\mathrm{V}}$ using the MW average $R_{\mathrm{V}}=3.1$ for Bayestar19 and the mean adopted by \citet{Zucker2025} $R_{\mathrm{V}}=3.32$ for DECaPS. Six of the Cepheid fields were covered by Bayestar19 and nine by DECaPS. Only the region of our closest Cepheid $\beta$~Dor is not featured in either of the maps due to its large negative Galactic latitude ($b=-32.8\degree$). The G-Tomo 3D dust map \citep{Lallement2022} gives $A_{\mathrm{V}}=0.06~$mag for $\beta$~Dor, and since its field stars are mostly more distant, we selected a slightly higher initial value of $A_{\mathrm{V}}=0.1$ for them.

Similarly, we queried the initial values for total-to-selective extinction ratio $R_{\mathrm{V}}$ from the 3D dust extinction curve map by \citet{Zhang2025}. The $R_{\mathrm{V}}$ map covered 10 of the 16 Cepheid fields. The extinction relation and the definition of $R_{\mathrm{V}}$ used by \citet{Zhang2025} differ from the ones in G23, which is why we let $R_{\mathrm{V}}$ vary freely and the final values are expectedly different from the initial values queried from the 3D map.
The five most distant Cepheid fields were out of reach of the map, and for their field stars, we chose the initial value $R_{\mathrm{V}}=3.1$ when the initial extinction came from the Baystar19, and $R_{\mathrm{V}}=3.32$ when from the DECaPS 3D dust map. For the field of $\beta$ Dor, which was not covered by the 3D $R_{\mathrm{V}}$ map either, we set $R_{\mathrm{V}}$ to the diffuse Galactic average $3.1$ with fixed $\pm0.1$ uncertainties. It was the only Cepheid for which we fixed the $R_{\mathrm{V}}$ of its field stars due to the already low number of stars with atmospheric parameters successfully determined from the spectra. 
The choice of $R_{\mathrm{V}}=3.1$ is justified by the fact that $\beta$ Dor is located far from the dusty Galactic plane and is therefore expected to be in the diffuse ISM domain.

We estimated the initial $\theta_{\mathrm{LD}}$ from a recent SBCR calibrated by \citet{Kiman2024} with the Gaia DR3 $G$ magnitudes and the $BP-RP$ color. According to Eqs. 5 and 10 and the coefficients in the first row of Table 1 in \citet{Kiman2024}, the angular diameter is
\begin{align}
    \theta_{\mathrm{LD}} &= \sqrt{10^{P - 0.4G_0}}\text{, where} \\
    P &= (-0.143X^2+1.22X+0.92) \cdot (1-0.017\cdot[Fe/H]),
\end{align}
where $X=BP_0-RP_0$ and $G_0$ are the dereddened color index and $G$ magnitude of the star. $[Fe/H]$ is the iron abundance that we approximated using the metal abundance $\left[\mathrm{M}/\mathrm{H}\right]$ determined from the spectra.
Based on their comparison with measured angular diameters, \citet{Kiman2024} reached an accuracy of $4\%$. However, the relation is calibrated for FGK and M dwarfs, while our sample includes more massive stars, which is why we use it merely for initial $\theta_{\mathrm{LD}}$ values.

Since we are using broadband photometry from 2MASS and WISE, we integrated the model SED over the filter transmission curves. For photon-counting detectors, the band-integrated model flux density was computed as
    \begin{align}
        F_{\mathrm{m, band}} = \frac{\int F_{\mathrm{m}}(\lambda)T(\lambda)\lambda d\lambda}{\int T(\lambda)\lambda d\lambda},
    \end{align}
where $F_{\mathrm{m}}(\lambda)$ is the flux density of the model SED and $T(\lambda)$ is the transmission curve of the filter. The resulting synthetic fluxes were compared with the observed flux densities. 

The success of each SED fit was evaluated with the reduced $\chi^2$ (or $\chi^2_{\nu}$) that we calculated after the fit using the conventional, unweighted $\chi^2_{\nu}$. We set the maximum $\chi^2_{\nu}$ limit to 20, and the 46 field stars that had a value higher than this were omitted from the determination of the Cepheid extinction. To ensure physically plausible values, we set a limit of $A_{\mathrm{V}} >0\,$mag for the extinction derived from the SED fits.  Furthermore, we added sigma clipping with a threshold of $5\sigma$ to exclude extinction values that diverged significantly from the rest of the field. For 274 stars, the value of $R_{\mathrm{V}}$ met the lower limit (2.3) of the G23 extinction relation, and for 27 stars the upper limit (5.6) indicating that $R_{\mathrm{V}}$ was poorly constrained, and we excluded these stars. Finally, $1230$ stars were used in 3D interpolation with an average of $77$ stars per Cepheid field.

\subsection{Cepheid extinction from 3D interpolation}

As the last step of the analysis, we estimated the Cepheid extinction. The distribution of field stars in space and their $A_{\mathrm{V}}$ values from the SED fitting compose a local three-dimensional distribution of extinction around the Cepheid. To estimate the extinction for the Cepheid, we interpolated in space to its position. The uncertainties in the field star $A_{\mathrm{V}}$, as well as the error bars of the Cepheid and field star parallax were propagated to the final Cepheid $A_{\mathrm{V}}$.

\begin{figure} 
\centering
\includegraphics[width=8.8cm]{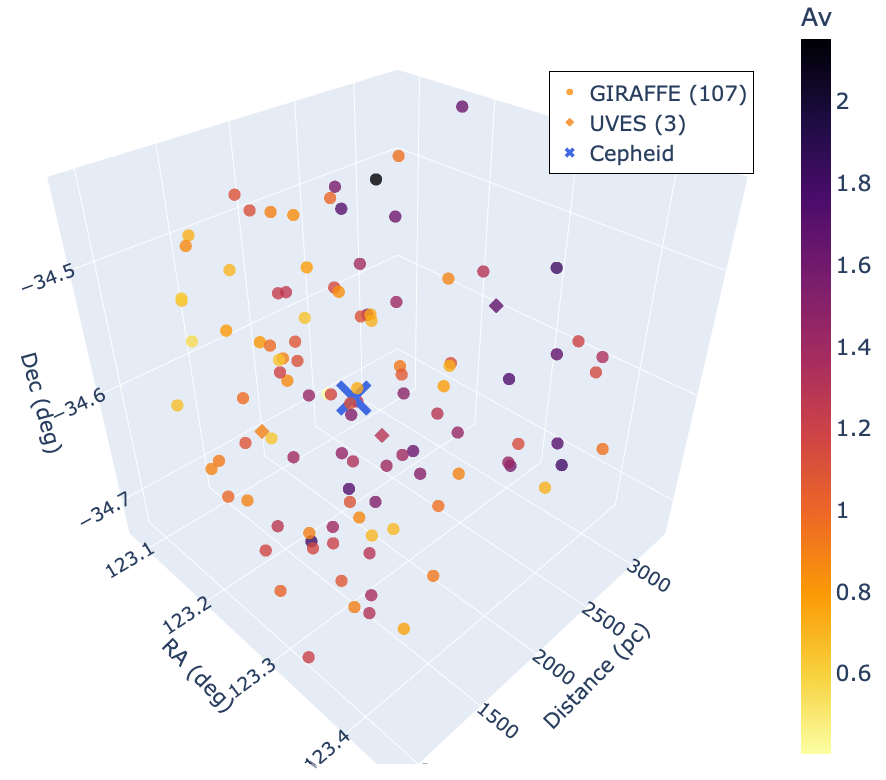}
   \caption{3D distribution of extinction around the Cepheid RS Pup. The Distance axis is greatly elongated compared to the RA and Dec axes. The stars observed with GIRAFFE and UVES spectrographs are shown with different symbols.}
      \label{fig: 3D plot}
\end{figure}

The resulting $A_{\mathrm{V}}$ grid is irregular and strongly elongated in the line-of-sight (LOS) direction. The distribution of field stars in the plane of the sky (POS) is dictated by the field of view of the FLAMES spectrograph, $25'$ in diameter. For example, RS Pup has a distance of approximately $1800\,$pc, meaning that at that distance the selected field stars are within $6.5\,$pc of the Cepheid in POS, while the LOS separation reaches $1500\,$pc (Fig. \ref{fig: 3D plot}). Furthermore, for nearby Cepheids like $\beta$ Dor and $\ell$ Car, the distribution of field stars is unbalanced, with most being more distant than the Cepheid. Analogously, for distant Cepheids like LS Pup and VZ Pup, they are mostly closer field stars. Interactive 3D representations of the extinction distribution around each Cepheid are given in the supplementary material.

We converted the star positions (RA, DEC, Plx) from Gaia DR3 to local Cartesian coordinates ($x$, $y$, $z$) centered at the Cepheid using a gnomonic projection and normalized them. In this way, we could calculate the field star separation from the Cepheid in parsecs:
\begin{align}
 r_i &= \sqrt{(x_i - x_{\mathrm{Cep}})^2 + (y_i - y_{\mathrm{Cep}})^2 + (z_i - z_{\mathrm{Cep}})^2}
\end{align}

\begin{figure} 
\centering
\includegraphics[width=8.8cm]{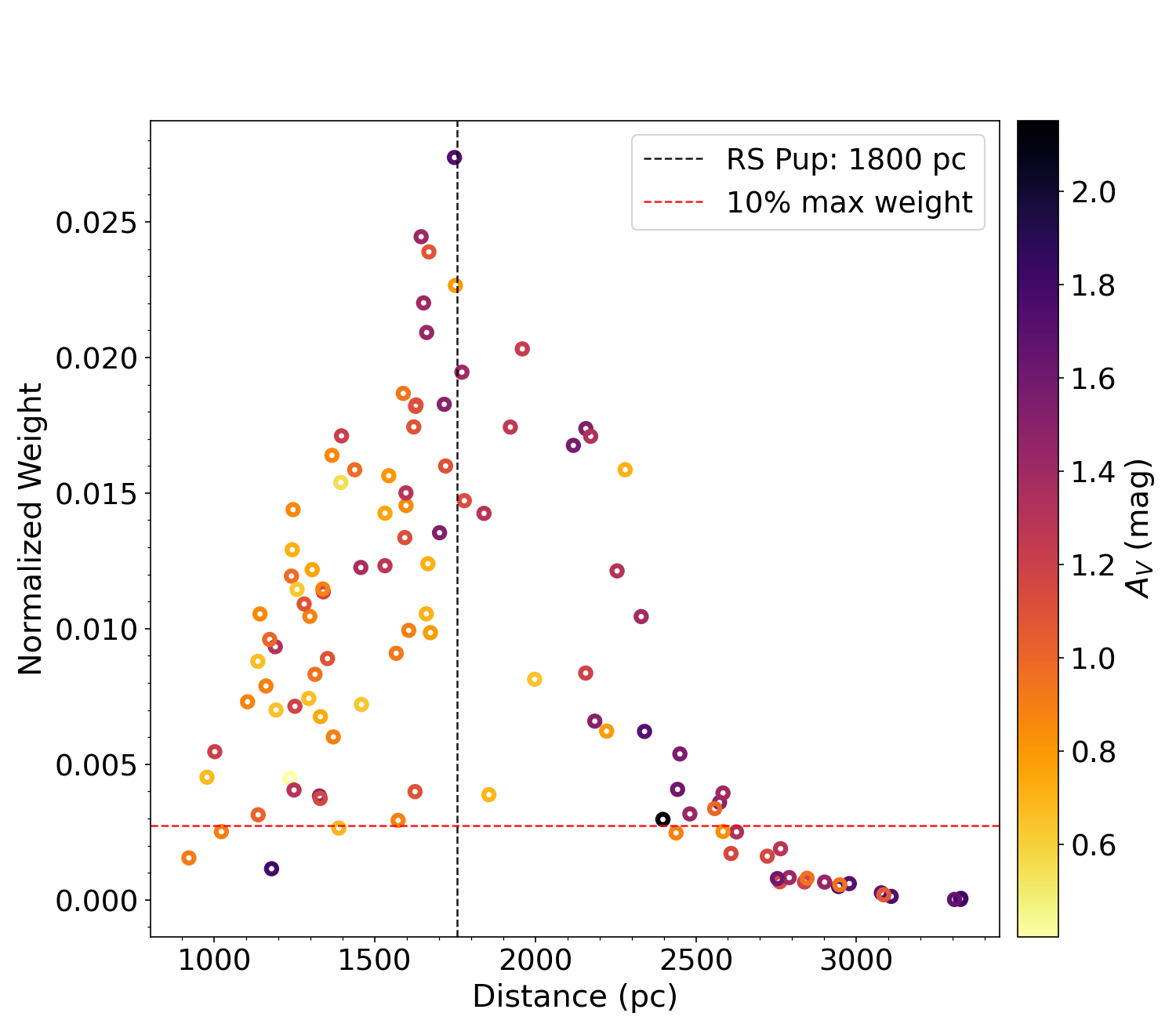}
   \caption{Weights set on the RS Pup field star extinction in the interpolation as a function of distance. The $A_{\mathrm{V}}$ values are shown with the colorbar. The black dashed line shows the RS Pup distance, and the red one where the weight is $10\%$ of the highest weight.}
      \label{fig: weights}
\end{figure}

The extinction at the Cepheid position $\textbf{x}_{\mathrm{Cep}}$ was estimated as the locally weighted average over the field star extinction $A_{\mathrm{V},i}$.
\begin{align}
    \hat{A}_{\mathrm{V}}(\textbf{x}_{\mathrm{Cep}}) = \frac{\sum_i w_i(r_i) A_{\mathrm{V},i}}{\sum_i w_i(r_i)} \label{Eq: mean Av}
\end{align}
We used a Gaussian kernel for the spatial weights \citep[Nadaraya–Watson kernel regression,][]{Nadaraya1964, Watson1964} based on the separation from Cepheid, with closer stars having greater impact than more distant ones. Each $A_{\mathrm V}$ value was further weighted by the inverse of its variance, thus emphasizing the SED fits with higher confidence for extinction.
\begin{align}
 w_i &= \exp\left(-\frac{r_i^2}{2h^2}\right) \cdot \frac{1}{\sigma_{A_{\mathrm{V},i}}^2} \label{Eq: weights}
\end{align}
The correlation length $h$ in the Gaussian spatial kernel characterizes how much of an impact the more distant field stars have on the Cepheid extinction. We set an individual $h$ value for each field as the median of the field star separation from the Cepheid. Typically, this meant $h\sim200-500\,$pc with higher values in sparser fields. Fig. \ref{fig: weights} shows the normalized weights as a function of distance from us for the RS Pup field with $h=459\,$pc

The Cepheid extinction uncertainties were estimated with resampling that combined both bootstrapping and Monte Carlo (MC) methods. Scatter in the field $A_{\mathrm{V}}$ values is expected, as the ISM is not smoothly distributed, but rather includes small-scale structures, e.g., clumps or small clouds, as well as variation in the dust properties. To account for this scatter, we used the bootstrapping method to randomly sample the field stars with replacement 300 times for each Cepheid field. To each of those resamples we applied the MC method, in which we sampled the $A_{\mathrm{V},i}$ and parallaxes of both the field stars and the Cepheid 300 times assuming a normal distribution for both variables with $\sigma_{A_{\mathrm{V},i}}$ and $\sigma_{\mathrm{Plx},i}$ from the SED fitting and the Gaia DR3 catalog, respectively. For each of these $300^2=90~000$ samples, we recalculated the ($x$, $y$, $z$) and the Cepheid extinction as a weighted average. We took the final extinction value for Cepheid as the mean of the sampled $A_{\mathrm{V}}$ values and the $1\sigma$ confidence level as the $16^{\mathrm{th}}$ and $84^{\mathrm{th}}$ percentiles.

\begin{figure} 
\centering
\includegraphics[width=8.8cm]{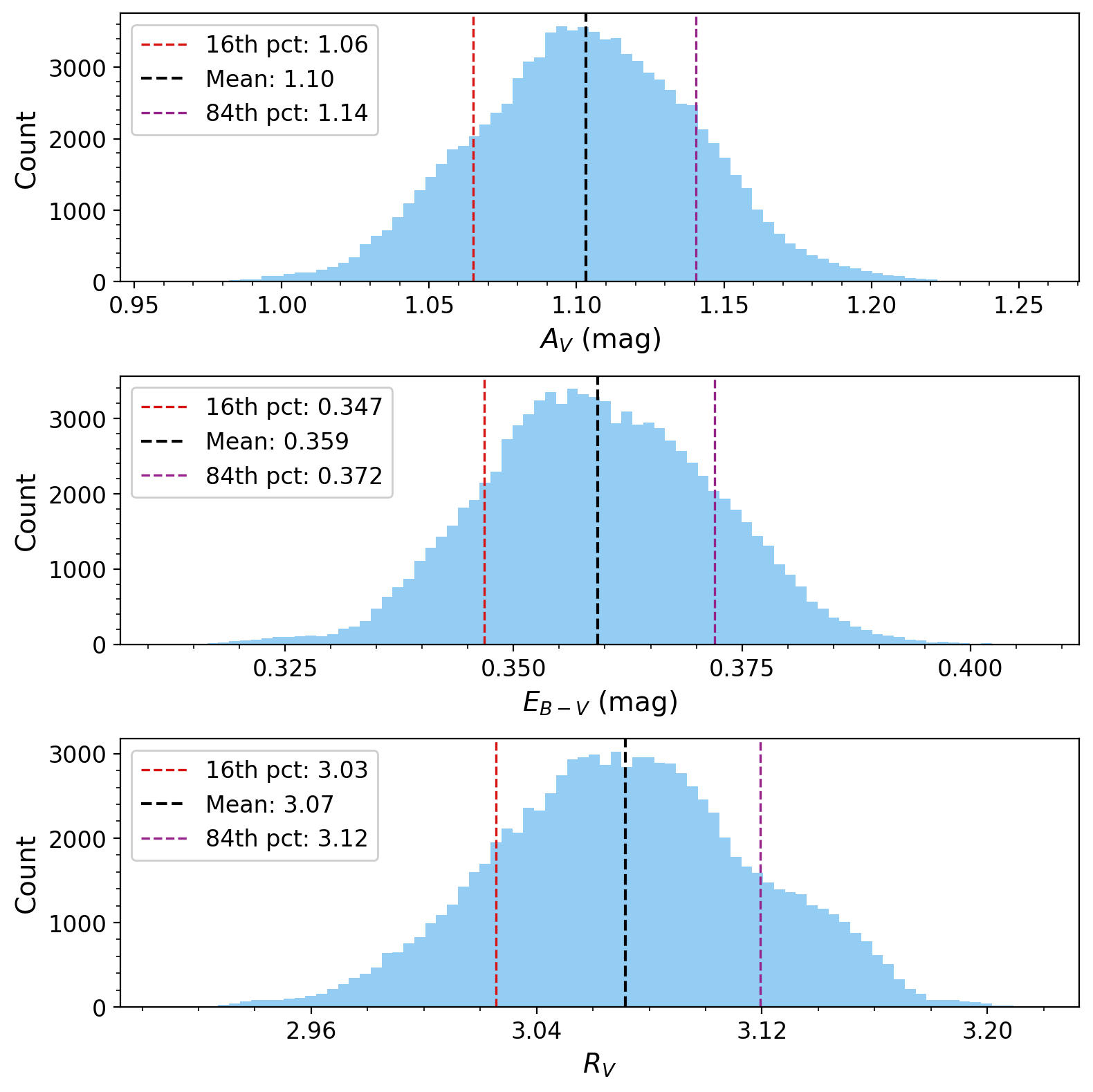}
   \caption{The sample distributions of the interpolated $A_{\mathrm{V}}$, $R_{\mathrm{V}}$, and $E_{\mathrm{B-V}}$ for RS Pup. The dashed lines show the mean and the 1$\sigma$ confidence level.}
      \label{fig: MC histo}
\end{figure}

We estimated the Cepheid $E_{\mathrm{B-V}}$ in a similar way to $A_{\mathrm{V}}$ in Eq. \ref{Eq: mean Av} and \ref{Eq: weights}. We opted to interpolate $A_{\mathrm{V}}$ and $E_{\mathrm{B-V}}$, as they are expected to vary more smoothly in space than their ratio $R_{\mathrm{V}}$. For this purpose, we calculated the individual field star color excess and its uncertainty based on the values found in the SED fits:
\begin{align}
    E_{\mathrm{B-V},i} &= \frac{A_{\mathrm{V,}i}}{R_{\mathrm{V,}i}} \\
    \sigma_{E_{\mathrm{B-V,}i}} &= \Bigg( \frac{\sigma_{A_{\mathrm{V,}i}}^2}{R_{{\mathrm{V,}i}}^2} + \frac{A_{{\mathrm{V,}i}}^2\sigma_{R_{\mathrm{V,}i}}^2}{R_{{\mathrm{V,}i}}^4} - \frac{2A_{{\mathrm{V,}i}}}{R_{{\mathrm{V,}i}}^3}\,\mathrm{Cov}(A_{{\mathrm{V,}i}}, R_{{\mathrm{V,}i}}) \Bigg)^{1/2},
\end{align}
where $\sigma_{A_{\mathrm{V,}i}}$ and $\sigma_{R_{\mathrm{V,}i}}$ are the respective uncertainties of $A_{\mathrm{V,}i}$ and $R_{\mathrm{V,}i}$, and $\mathrm{Cov}(A_{{\mathrm{V,}i}}, R_{{\mathrm{V,}i}})$ is their covariance. Following the interpolation, we calculated $R_{\mathrm{V}} = A_{\mathrm{V}}/E_{\mathrm{B-V}}$ for each MC sample. In this way, we obtained the $1\sigma$ confidence level likewise for $R_{\mathrm{V}}$ directly from the sampling. Fig. \ref{fig: MC histo} shows the distributions of the RS Pup $A_{\mathrm{V}}$, $E_{\mathrm{B-V}}$, and $R_{\mathrm{V}}$ samples with the mean values and confidence intervals.

\section{Results}

\begin{figure*}
\centering
\includegraphics[width=18cm]{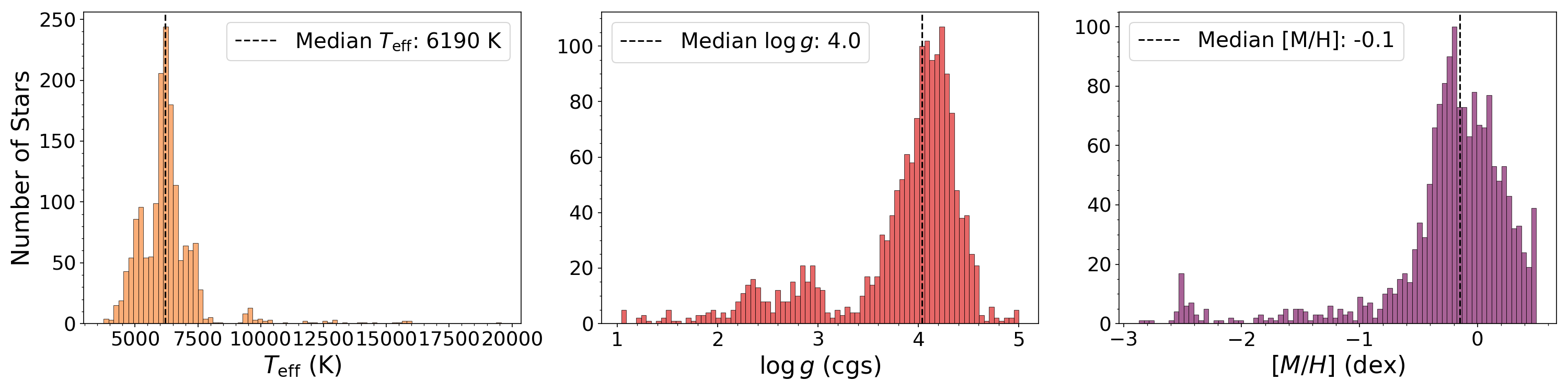}
  \caption{Distributions of the effective temperature ($T_{\mathrm{eff}}$), surface gravity ($\log g$), and metal abundance ($\left[\mathrm{M}/\mathrm{H}\right]$) found in the spectral fitting for the field stars of the 16 Cepheids.}
     \label{fig: param distributions}
\end{figure*}

\begin{figure*}
\centering
\includegraphics[width=18cm]{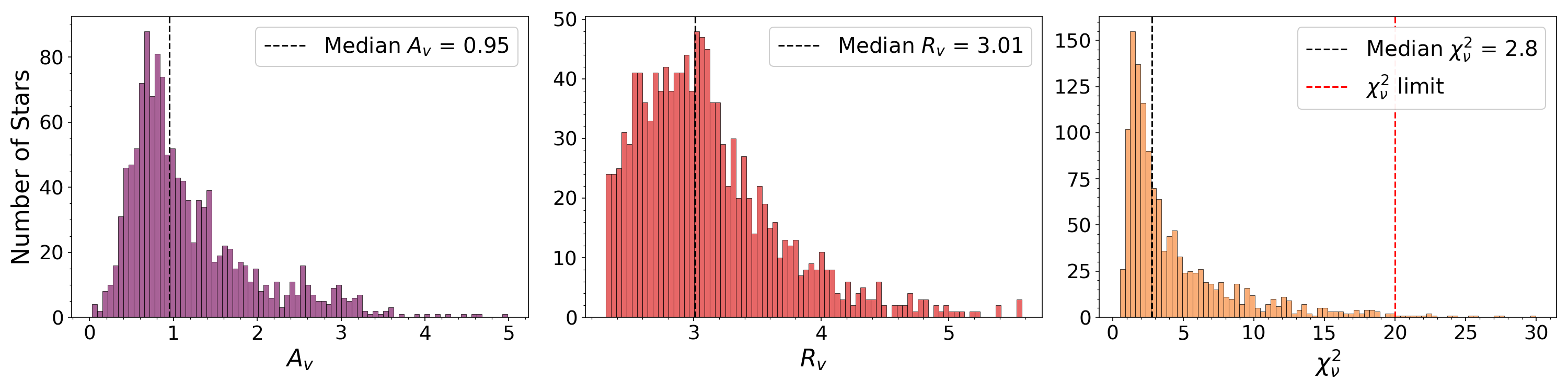}
  \caption{Distributions of the extinction ($A_{\mathrm{V}}$), total-to-selective extinction ratio ($R_{\mathrm{V}}$), and reduced chi-squared ($\chi^2_{\nu}$) found in the SED fitting for the field stars of the 16 Cepheids.}
     \label{fig: Av_Rv_chi2 distributions}
\end{figure*}

\subsection{Field star atmospheric parameters}
We derived the atmospheric parameters of the field stars from the MIST isochrones ($\log g$ for stars observed with GIRAFFE) and by fitting the stellar spectra with ATLAS9 atmosphere models ($T_{\mathrm{eff}}$ and $\left[\mathrm{M}/\mathrm{H}\right]$ for all stars and $\log g$ for stars observed with UVES). The distributions of the parameters with the $1609$ field stars, for which we could successfully determine the parameters from the spectra, are shown in Fig. \ref{fig: param distributions}. The values for $T_{\mathrm{eff}}$ are between $3750$ and $19600\,$K with a median of $6190\,$K, for $\log g$ between $1.04-5.0$ with a median of $4.0$, and for $\left[\mathrm{M}/\mathrm{H}\right]$ between $(-2.87)-0.5$ with a median of $-0.1$. The majority of stars are centered around their medians for all three parameters, indicating that main-sequence stars are dominant, with a minority of giants with $\log g \lesssim 3.5$. All parameters found for field stars, including [$\alpha$/Fe], $V_{\mathrm{mic}}$, $V_{\mathrm{mac}}$, and $\upsilon \sin{i}$, are provided in the supplementary material.

\subsection{Field star extinction}
The extinction for each field star was estimated by fitting its SED with the ATLAS9 models and the extinction relation by \citet{Gordon2023}. In Fig. \ref{fig: Av_Rv_chi2 distributions}, we show distributions of the resulting $A_{\mathrm{V}}$ and $R_{\mathrm{V}}$, and the $\chi^2_{\nu}$ of the field stars that were finally used to derive the Cepheid extinction. $A_{\mathrm{V}}$ found values between $0.03-4.99\,$mag with a median of $0.95\,$mag, highly dependent on the distance of the star. $R_{\mathrm{V}}$ varied within $2.31-5.57$, with a median of $3.01$. For many stars we found low $R_{\mathrm{V}}$ values, but the median is still close to the MW average for the diffuse medium $3.1$.
The median $\chi^2_{\nu}$ was $2.8$, at its lowest it was $0.3$, and the highest accepted value was set to $20$. 

The limb-darkened angular diameter was likewise fitted with its initial values calculated with the SBCR by \citet{Kiman2024}. The $\theta_{\mathrm{LD}}$ values we found are on average $10\%$ higher than these initial values. We found a strong positive correlation ($r\sim0.8$) between $\theta_{\mathrm{LD}}$ and $R_{\mathrm{V}}$ in the SED fitting and recommend taking this into account if adopting their values in other studies. Between $A_{\mathrm{V}}$ and $R_{\mathrm{V}}$ the correlation was moderate ($r\sim0.4-0.5$). More details on parameter correlations are given in the Appendix~\ref{appendix: correlations}.
Parameter values found in the SED fitting for individual field stars are also given in the supplementary material.

\subsection{Cepheid $A_{\mathrm{V}}$, $E_{\mathrm{B-V}}$, and $R_{\mathrm{V}}$}

\begin{table*}
\caption{Extinction for the 16 Cepheids of our pilot sample.}
\label{tab:cepheids}
\centering
\renewcommand{\arraystretch}{1.4}
\begin{tabular}{lccccccccc}
\hline\hline 
Cepheid & RA (deg) & Dec (deg) & Plx (mas) & Period (d) & RUWE & $\mathcal{N}_{\mathrm{star}}$ & $A_V$ (mag) & $R_V$ & $E_{\mathrm{B-V}}$ (mag) \\
\hline
$\ell$~Car   & 146.312 & -62.508 & $1.98 \pm 0.11$ & 35.54 & 2.39 & 80 & $0.563_{-0.020}^{+0.020}$ & $3.048_{-0.067}^{+0.071}$ & $0.185_{-0.008}^{+0.008}$ \\
U~Car   & 164.451 & -59.732 & $0.55 \pm 0.02$ & 38.80 & 1.23 & 89 & $0.902_{-0.045}^{+0.045}$ & $3.123_{-0.089}^{+0.090}$ & $0.289_{-0.015}^{+0.015}$ \\
VY~Car  & 161.136 & -57.565 & $0.55 \pm 0.02$ & 18.91 & 0.92 & 78 & $0.688_{-0.033}^{+0.033}$ & $3.100_{-0.081}^{+0.081}$ & $0.222_{-0.011}^{+0.012}$ \\
KN~Cen  & 204.154 & -64.558 & $0.23 \pm 0.02$ & 34.05 & 1.03 & 88 & $2.696_{-0.069}^{+0.070}$ & $3.858_{-0.097}^{+0.096}$ & $0.699_{-0.034}^{+0.033}$ \\
S~Cru   & 193.592 & -58.431 & $1.32 \pm 0.02$ & 4.69  & 0.94 & 87 & $0.775_{-0.023}^{+0.022}$ & $2.960_{-0.054}^{+0.055}$ & $0.262_{-0.007}^{+0.007}$ \\
$\beta$~Dor & 83.406  & -62.490 & $2.930 \pm 0.14$ & 9.84  & 4.54 & 47 & $0.368_{-0.032}^{+0.033}$ & $3.764_{-0.235}^{+0.237}$ & $0.098_{-0.012}^{+0.012}$ \\
CV~Mon  & 99.270  & +3.064  & $0.57 \pm 0.01$ & 5.38  & 1.10 & 80 & $1.790_{-0.113}^{+0.119}$ & $3.200_{-0.081}^{+0.083}$ & $0.599_{-0.032}^{+0.031}$ \\
T~Mon   & 96.304  & +7.086  & $0.71 \pm 0.05$ & 27.02 & 1.72 & 75 & $0.812_{-0.050}^{+0.050}$ & $3.130_{-0.054}^{+0.054}$ & $0.260_{-0.015}^{+0.015}$ \\
S~Mus   & 183.196 & -70.152 & $1.17 \pm 0.09$ & 9.66  & 4.50 & 87 & $0.698_{-0.023}^{+0.023}$ & $3.333_{-0.064}^{+0.069}$ & $0.210_{-0.008}^{+0.008}$ \\
LS~Pup  & 119.747 & -29.308 & $0.19 \pm 0.02$ & 14.15 & 1.25 & 60 & $1.506_{-0.067}^{+0.062}$ & $3.021_{-0.088}^{+0.087}$ & $0.499_{-0.021}^{+0.022}$ \\
RS~Pup  & 123.268 & -34.579 & $0.57 \pm 0.02$ & 41.39 & 1.16 & 110 & $1.104_{-0.038}^{+0.040}$ & $3.070_{-0.042}^{+0.045}$ & $0.360_{-0.012}^{+0.012}$ \\
VZ~Pup  & 114.647 & -28.500 & $0.20 \pm 0.01$ & 23.17 & 1.24 & 93 & $1.606_{-0.037}^{+0.039}$ & $3.066_{-0.058}^{+0.060}$ & $0.524_{-0.016}^{+0.016}$ \\
WX~Pup  & 115.496 & -25.876 & $0.37 \pm 0.02$ & 8.94  & 1.06 & 78 & $0.717_{-0.043}^{+0.042}$ & $3.119_{-0.079}^{+0.075}$ & $0.230_{-0.012}^{+0.012}$ \\
X~Pup   & 113.196 & -20.910 & $0.38 \pm 0.02$ & 25.96 & 1.04 & 60 & $0.946_{-0.071}^{+0.071}$ & $2.886_{-0.083}^{+0.084}$ & $0.328_{-0.023}^{+0.023}$ \\
CS~Vel  & 145.293 & -53.816 & $0.26 \pm 0.01$ & 5.90  & 0.91 & 69 & $2.138_{-0.119}^{+0.120}$ & $3.556_{-0.121}^{+0.124}$ & $0.601_{-0.037}^{+0.036}$ \\
RZ~Vel  & 129.255 & -44.115 & $0.65 \pm 0.02$ & 20.40 & 1.24 & 96 & $1.021_{-0.040}^{+0.038}$ & $2.846_{-0.051}^{+0.046}$ & $0.359_{-0.013}^{+0.013}$ \\
\hline
\end{tabular}
\end{table*}

As the principal result of our study, we provide accurate extinction values with $1\sigma$ uncertainties for the 16 Galactic Cepheids, given in Table \ref{tab:cepheids} alphabetically ordered by constellation. Detailed in Table \ref{tab:cepheids} is also the number of field stars ($\mathcal{N}_{\mathrm{star}}$) used in the interpolation for each Cepheid. The relative one-sided error is between $2-9\%$ with an average of $5\%$. The most significant contribution to the error budget comes from the scatter of $A_{\mathrm{V}}$ in the field. For example, we found $A_{\mathrm{V}}=1.79_{-0.12}^{+0.11}\,$mag for the Cepheid CV~Mon, which has considerable variation in the field extinction: just within $100\,$pc from the Cepheid there are 13 stars, whose $A_{\mathrm{V}}$ varies from $1.5$ to $4.6\,$mag, strongly depending on the line of sight. If we do not include the bootstrapping that factors in this variation in the uncertainty of the Cepheid $A_{\mathrm{V}}$, thus propagating only the field star extinction and parallax uncertainties, we find $A_{\mathrm{V}}=1.79_{-0.05}^{+0.05}\,$mag. This shows that in cases like CV Mon, the limited number of field stars and their irregular distribution in space are insufficient to resolve the small-scale structures in the ISM, which is reflected in the higher uncertainties.

Furthermore, we provide estimates for the color excess $E_{\mathrm{B-V}}$, and, unlike most previous extinction studies, for the total-to-selective extinction ratio $R_{\mathrm{V}}$ for each Cepheid (Table \ref{tab:cepheids}). Within our small sample, we find variation in Cepheid $R_{\mathrm{V}}$ from $2.85$ to $3.86$, demonstrating that the MW average $R_{\mathrm{V}}=3.1$ would be insufficient to model extinction in all fields. The mean one-sided statistical uncertainty for $R_{\mathrm{V}}$ is only $2.6\%$, which is partly explained by the smaller variation in its value compared to $A_{\mathrm{V}}$. It should be noted that the quoted precision in $R_{\mathrm{V}}$ is conditional on the adopted \mbox{ATLAS9} atmosphere models \citep{Castelli2003} and the \citet{Gordon2023} extinction relation.
We tested interpolating $R_{\mathrm{V}}$ instead of $E_{\mathrm{B-V}}$, which resulted in on average $5\%$ lower values for Cepheid $R_{\mathrm{V}}$ and 
a mean statistical uncertainty of $1.7\%$. Interpolating $E_{\mathrm{B-V}}$ instead yields $R_{\mathrm{V}}$ values more consistent with the diffuse ISM average and larger, more representative uncertainties, which is why we adopted it as our method. The choice between interpolating $E_{\mathrm{B-V}}$ or $R_{\mathrm{V}}$ has no impact on the interpolation of $A_{\mathrm{V}}$.

\subsection{Comparison to literature values} \label{sec: literature comparison}

\begin{figure*}
\centering
\includegraphics[width=18cm]{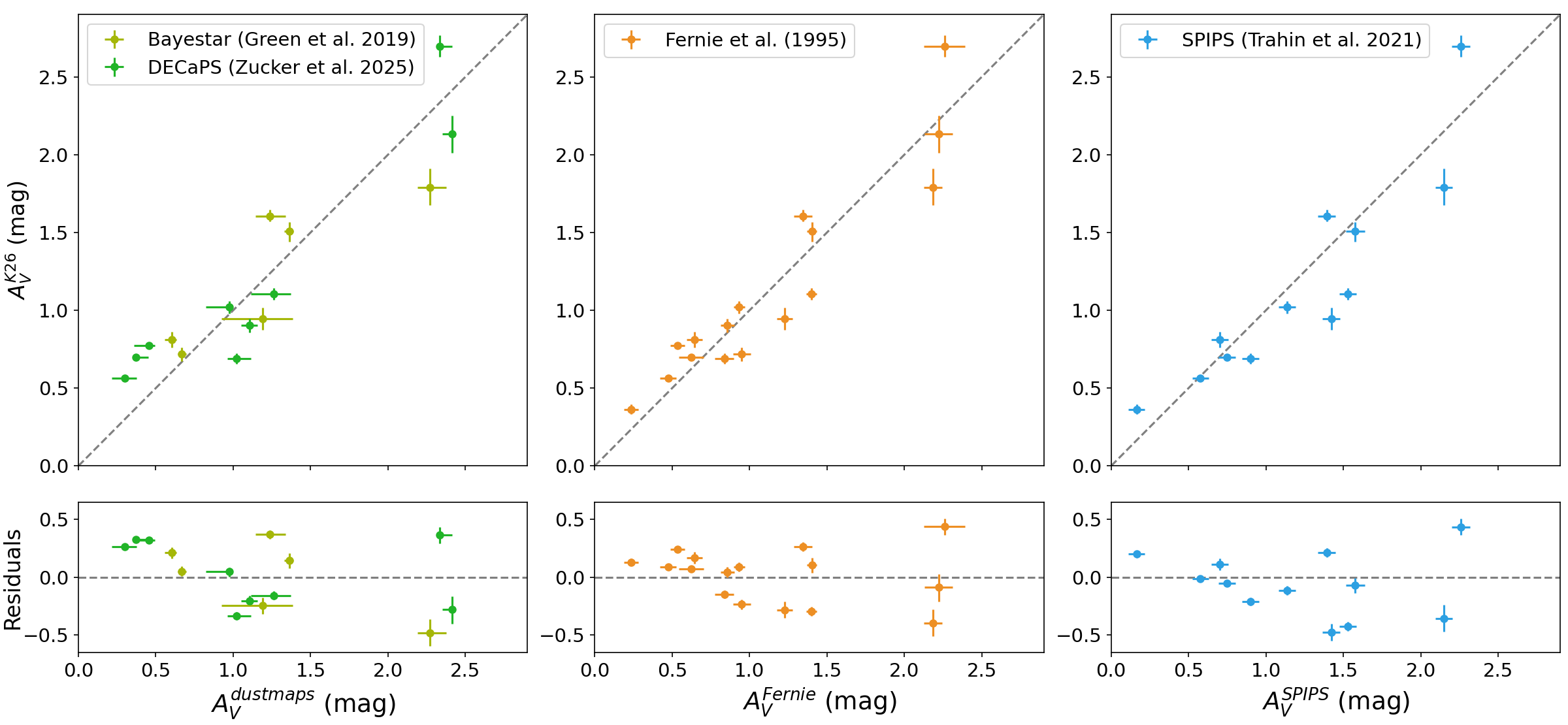}
  \caption{Comparison of our Cepheid extinction results (y-axis, K26) to literature (x-axis). All of the literature values have been converted from $E_{\mathrm{B-V}}$ to $A_{\mathrm{V}}$, and the values from \citet{Fernie1995} and \citet{Trahin2021} have been scaled (see Sec. \ref{sec: literature comparison} for details).}
     \label{fig: comparison}
\end{figure*}

We compared our results with the Cepheid extinction in the literature (Fig. \ref{fig: comparison}). The first two references are the 3D dust maps Bayestar19 \citep{Green2019} and DECaPS \citep{Zucker2025}, the same ones we used for initial $A_{\mathrm{V}}$ for the field stars in the SED fitting (Sec. \ref{sec: SEDs}). The angular resolution of Bayestar19 varies across the map, but \citet{Green2019} estimate a typical resolution of $6.8'$, while \cite{Zucker2025} quote $1'$ angular resolution for DECaPS 3D dust map. In the line-of-sight direction, the distance modulus from $\mu=4$ to $19\,$mag is evenly discretized into 120 bins, each with a size of $0.125\,$mag in both 3D maps. Together, they cover the entire Galactic plane ($|b|<10$), providing reddening estimates for all 16 Cepheids, except for $\beta$ Dor.
We queried the mean $E_{\mathrm{B-V}}$ and $1\sigma$ uncertainties with the Cepheid coordinates and distances (Gaia DR3 parallaxes), and converted to $A_{\mathrm{V}}$ with $R_{\mathrm{V}}=3.1$ for Bayestar19 and $3.32$ for DECaPS 3D dust map. The mean difference between the dust map $A_{\mathrm{V}}$ and ours is small, ours being 0.02 mag higher.
Their relative one-sided errors vary significantly from $2$ to $22\%$ (average $7\%$) for Bayestar19 and from $0.1$ to $28\%$ (average $10\%$) for DECaPS. The wide range and unrealistically low error bars for some Cepheids question the accuracy of these extinction results.

The second set of reference values comes from the database of 505 Galactic Cepheids compiled by \citet{Fernie1995}. The database is a compilation of the major color excess determinations published between 1975 and 1995. We included it in the comparison, since its reddening values are widely used to this day in the Cepheid field \citep[e.g.,][]{Benedict2007, Gaia2017, Narloch2023, Bailleul2025}. The database includes mean $E_{\mathrm{B-V}}$ values with related errors for all 16 Cepheids. The relative error for these values is within $2-20\%$ with an average of $7\%$.
A systematic trend in \citet{Fernie1995} $E_{\mathrm{B-V}}$ was first suggested by \citet{Tammann2003}, and later by \citet{Groenewegen2018}, who used a scaling term of 0.94, and which we also adopted here. The $E_{\mathrm{B-V}}$ was converted to $A_{\mathrm{V}}$ using $R_{\mathrm{V}}=3.1$. Our $A_{\mathrm{V}}$ values are close, being on average only 0.01 mag higher than the scaled values by \citet{Fernie1995}.

We also considered the results of \citet{Trahin2021} using the SPIPS algorithm \citep{Merand2015} for a comprehensive modeling of 63 Galactic Cepheids, of which 12 are in our sample. The reddening in SPIPS is parametrized with the color excess and uses the extinction relation by \citet{Fitzpatrick1999} with $R_{\mathrm{V}}=3.1$. The $E_{\mathrm{B-V}}$ is fitted simultaneously with multiple parameters constraining, i.a, the size of the star, the pulsation period, and IR excess. We converted the $E_{\mathrm{B-V}}$ values provided for 12 of our Cepheids in \citet{Trahin2021} into $A_{\mathrm{V}}$ using a function provided in  SPIPS\footnote{\url{https://github.com/amerand/SPIPS}: Alambda\_Exctinction() in spips.py} giving also $T_{\mathrm{eff}}$ of the Cepheid as input. The relative uncertainties for the 12 values are within $3-33\%$ with an average of $7\%$.
As SPIPS is designed for the modeling of pulsating stars \citep[primarily for Cepheids, but it has also been applied to RR Lyrae, see][]{Bras2024}, it adopts a reddening definition that is optimized for stars with $T_{\mathrm{eff}}=4500-6500\,$K and that differs from the standard one used in the extinction relations. As a result, the extinction values derived with SPIPS modeling tend to be higher than in the literature. To make them comparable to our results, we divided them by the scaling factor $1.16$ following \citet{Bras2026}. 
Additionally, SPIPS includes modeling of IR emission above $1.2~\mu$m, contributed to a circumstellar envelope (CSE), but assumes no emission or absorption by the CSE below this wavelength (See Sec. \ref{sec: CSM discussion}). In case of a dusty CSE, there would also be absorption in the UV and visible domains, and SPIPS could increase the extinction to compensate for it.
Even after scaling, the $A_{\mathrm{V}}$ extinction by \citet{Trahin2021} is on average 0.07 mag higher than ours.

There is agreement between our and these Cepheid extinction literature values (Fig. \ref{fig: comparison}, Cepheid by Cepheid comparison in Fig. \ref{fig: histo_comparison}). The closest to our results are the extinction values from the \citet{Fernie1995}, and the relative differences do not suggest any obvious bias compared to the literature sources. We found both lower and higher values for the 16 Cepheids, with considerably higher extinction for S~Cru, VZ~Pup, and KN~Cen than any of the references (Fig. \ref{fig: histo_comparison}). On average, we achieved the highest precision, especially for lower extinction values.

\section{Discussion} \label{sec: discussion}

\subsection{Implications on the field}
This approach of determining Cepheid extinction using nearby field stars shows great potential in improving the calibration of the Galactic PL relation and the Cepheid distance determination with the PoP method. Our low $A_{\mathrm{V}}$ uncertainties will reduce the overall PL relation error budget, and improved accuracy is expected to reduce the scatter, especially at optical wavelengths where the impact of extinction is greater. Based on the results of this pilot study, we did not find a significant offset between our values and the literature \citep{Fernie1995, Green2019, Zucker2025}, but the sample of 16 Cepheids is too small to detect a potential large-scale bias in existing reddening estimations and their impact on the PL relation. Our full sample of 100 Galactic Cepheids will provide more definitive answers in upcoming studies.

Similarly, the improved extinction values can result in tighter Cepheid SBCR calibrations reducing uncertainties related to both their slopes and zero-points. The lower systematics in the SBCRs will then be reflected in smaller uncertainties of Cepheid distances when applying them in the PoP method. Although the impact of extinction is lesser in the $V-K$ color \citep{Nardetto2023}, which is typically used, the effect will be more notable when aiming for a multi-chromatic empirical adaptation of the PoP method.
Beyond these two applications, our results enable straightforward and precise extinction corrections in other Cepheid-related studies.
Going even further, the method could equally be applied to estimating the extinction of other variable stars, or almost any type of target in the Milky way, for that matter.

\subsection{Inconstant $R_{\mathrm{V}}$}
A considerable distinction between our work and most extinction studies is that we did not assume a constant shape of the extinction curve ($R_{\mathrm{V}}$) in the Milky Way. The $A_{\mathrm{V}}$ and $R_{\mathrm{V}}$ parameters are generally strongly correlated in the fitting of stellar SEDs, but we found that the introduction of the Gaia XP spectra mostly broke this degeneracy. However, we see a strong correlation between the parameters $R_{\mathrm{V}}$ and $\theta_{\mathrm{LD}}$ (Appendix \ref{appendix: correlations}).

Observations show large sightline-to-sightline variation in extinction due to different dust properties (grain composition, shape, and size distribution). Changes in $R_{\mathrm{V}}$ are linked to changes in grain size distribution: The common understanding has been that the higher $R_{\mathrm{V}}$ correlates with the denser regions and therefore with the larger average grain size \citep{Weingartner2001}. However, contrary to this perception, \citet{Zhang2025} found a U-shaped trend of $R_{\mathrm{V}}$ with density, so that the value of $R_{\mathrm{V}}$ decreases from diffuse to moderate-density ISM, and increases sharply in the densest regions. We used their 3D $R_{\mathrm{V}}$ map for initial values in the SED fitting of the ten Cepheid fields that it covered, which improved the precision of our results. Thus, we confirm that a constant $R_{\mathrm{V}}$ is generally insufficient to model extinction in various parts of the Milky Way.

\subsection{Circumstellar envelopes and small-scale ISM variations} \label{sec: CSM discussion}
Important aspects that we are mostly unable to factor in with this method are the contributions from CSEs and significant variations in dust density on scales smaller than the distances between field stars. Many Cepheids are speculated to have CSEs, but their actual presence, composition, and contribution to extinction are not well known for most Cepheids. A compact envelope was first resolved with VLTI/VINCI around $\ell$ Car by \citet{Kervella2006} followed by other interferometric observations reporting CSEs of more Cepheids \citep{Merand2006, Merand2007, Gallenne2013b, Hocde2021}.
Studying the origin of the IR excess in the SEDs of five Cepheids (including RS Pup), \citet{Hocde2020} found that it could not be explained by CSE models that included dust. Rather, they could reproduce the IR excess with a thin shell of ionized gas. The gaseous nature of CSEs was confirmed by \citet{Hocde2025b}, as they reported ALMA observations of ionized gas emission near $\ell$ Car. Nevertheless, more observations and modeling are required to draw any firm conclusions on the Cepheid CSEs. As their true nature remains uncertain, so does their contribution to extinction. However, given the moderate magnitude of the detected IR excess and the more likely gaseous rather than dusty envelopes, the contribution is expected to be small.

As Population I stars, Classical Cepheids are relatively young, with the long-period ones being more luminous and massive and therefore, even younger. This means that Cepheids can be embedded in star forming regions with highly inhomogeneous distribution of dust and gas. From our Cepheid sample, RS Pup is known to be located in a reflection nebula, in which small-scale dust density variations are likely. In such environments, our method becomes more sensitive to the distribution and distance of the field stars relative to the Cepheid. The lower extinction that we find for RS Pup compared to the literature (Fig. \ref{fig: histo_comparison}) could be due to these unresolved structures. Other Cepheids, for which the nearby regions are not as well characterized observationally, could be impacted by a similar effect (e.g., X Pup or CV Mon). Simultaneously, our method combined with other extinction studies introduces an opportunity to confirm or discard the disputed presence of CSEs and extended nebulae around Cepheids.

\section{Conclusions}
We have introduced a novel approach to estimate the impact of interstellar extinction on Galactic Cepheid variables. By estimating Cepheid extinction with three-dimensional interpolation from the extinction of field stars, we avert the complications and uncertainties related to the Cepheid pulsation. Our results for the pilot sample of 16 Cepheids with a comprehensive range of properties show the highest precision and no obvious bias in the extinction values compared to the literature. 

The field star extinction, which we parametrized with $A_{\mathrm{V}}$, was estimated in the fitting of the stellar SEDs. Instead of simultaneously deriving the atmospheric parameters, we determined them from the VLT/FLAMES spectra. In this way, we avoided the well-known degeneracy between effective temperature and extinction, which promotes the accuracy of the resulting $A_{\mathrm{V}}$. We also derived individual values for the $R_{\mathrm{V}}$ parameter (total-to-selective extinction ratio) and found that the inclusion of Gaia XP spectra in the SEDs effectively broke its degeneracy with $A_{\mathrm{V}}$. With our results, we confirm that $R_{\mathrm{V}}$ varies significantly within the Milky Way and that fixing it to the average $3.1$ would weaken the precision of the extinction modeling.

A challenge to this approach arises from small-scale structures in the ISM. The achieved precision depends on the length scale over which the dust properties and density vary and the distribution of field stars around the Cepheid. For environments of highly heterogeneous dust content and few stars, the uncertainties of the Cepheid extinction are inevitably larger. In addition, this method does not account for the contribution of potential circumstellar envelopes of Cepheids to extinction. This property, on the other hand, could be used in the search for the CSEs when compared with direct extinction measurements of the Cepheid. 
Extension of the method to a larger sample of 100 Cepheids is planned, and the derived extinction is expected to improve the calibration of the Galactic PL relation together with the upcoming fourth data release of the Gaia mission.

\medskip
\noindent \textbf{Data availability}
{\small Table \ref{tab:cepheids} and the Cepheid field star catalogs will be available at the CDS under designation J/A+A/XXX/YYY upon publication of this article. The raw and processed FLAMES data are available at \url{https://archive.eso.org/} under Program ID 112.25ES.001.}

\begin{acknowledgements}
      The research leading to these results has received funding from the European Research Council (ERC) under the European Union’s Horizon 2020 research and innovation program (project UniverScale, grant agreement 951549).
      We acknowledge support from the Polish-French Marie Skłodowska-Curie and Pierre Curie Science Prize awarded by the Foundation for Polish Science.
      MJ acknowledges the support of the Research Council of Finland Grant No. 348342.
      AG acknowledges support from the Agencia Nacional de Investigación y Desarrollo (ANID) through FONDECYT Regular grant 1241073.
      BP acknowledges support from the Polish National Science Center grant SONATA BIS 2020/38/E/ST9/00486.
      The authors also acknowledge the support of the French Agence Nationale de la Recherche (ANR), under grant ANR-23-CE31-0009-01 (Unlock-pfactor).
      This work has made use of data from the European Space Agency (ESA) mission Gaia (https://www.cosmos.esa.int/gaia), processed by the Gaia Data Processing and Analysis Consortium (DPAC, 
      \url{https://www.cosmos.esa.int/web/gaia/dpac/consortium}). Funding for the DPAC has been provided by national institutions, in particular the institutions participating in the Gaia Multilateral Agreement.
      This publication makes use of data products from the Two Micron All Sky Survey, which is a joint project of the University of Massachusetts and the Infrared Processing and Analysis Center/California Institute of Technology, funded by the National Aeronautics and Space Administration and the National Science Foundation.
      This publication makes use of data products from the Wide-field Infrared Survey Explorer, which is a joint project of the University of California, Los Angeles, and the Jet Propulsion Laboratory/California Institute of Technology, and NEOWISE, which is a project of the Jet Propulsion Laboratory/California Institute of Technology. WISE and NEOWISE are funded by the National Aeronautics and Space Administration.
      This work made use of Astropy (available at \url{http://www.astropy.org}), a community-developed core Python package and an ecosystem of tools and resources for astronomy \citep{Astropy2013, Astropy2018, Astropy2022}, the Numpy library \citep{harris2020}, the SciPy library \citep{Virtanen2020}, the astroquery library \citep{Ginsburg2019}, the Matplotlib graphics environment \citep{Hunter2007}, the synthetic photometry synphot package \citep{synphot2018}, and the dust\_extinction package \citep{Gordon2024}.
      This research has made use of the SIMBAD database and the VizieR catalogue access tool, CDS, Strasbourg Astronomical Observatory, France \cite{Ochsenbein2000}, and the SVO Filter Profile Service "Carlos Rodrigo", funded by MCIN/AEI/10.13039/501100011033/ through grant PID2023-146210NB-I00 \citep{Rodrigo2012, Rodrigo2024, Rodrigo2020}.
\end{acknowledgements}

\bibliographystyle{aa}
\bibliography{ref}

\onecolumn
\begin{appendix}
\section{UVES spectra from the VLT/FLAMES observations} \label{appendix: UVES spec}
\begin{figure*}[h]
    \includegraphics[width=18.5cm]{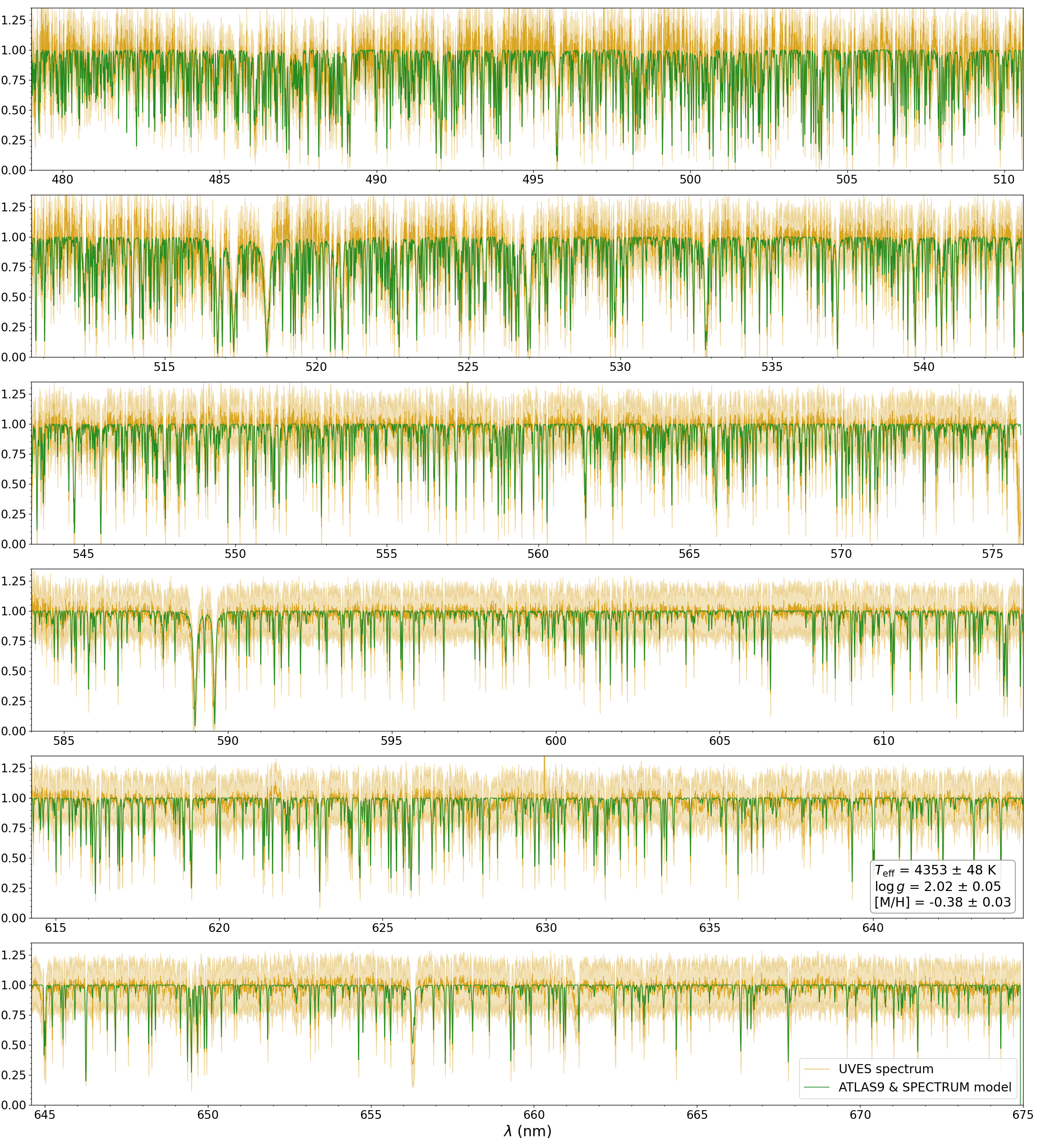}
    \caption{Observed (yellow) and fitted model spectrum (green) of RS Pup field star Gaia DR3 5546476858619276672. The light yellow shows uncertainties of the observed spectra estimated as the RMS of the continuum regions.}
    \label{fig: UVES spec}
\end{figure*}

The field stars observed with the UVES spectrograph ($R\sim47000$) were modeled with synthetic spectra built from ATLAS9 atmosphere models \citep{Castelli2003} with the radiative transfer code \texttt{SPECTRUM} \citep{Gray1994}. The observed and modeled spectrum for the RS Pup field star Gaia DR3 5546476858619276672 ($G=14.5$) is shown in Fig. \ref{fig: UVES spec}. A systematic uncertainty of $1\%$ has been added to the error bars of the atmospheric parameters.

\nolinenumbers
\begin{multicols}{2}

\section{Surface gravity from the UVES spectra vs. the MIST isochrones} \label{appendix: logg}
We compared the $\log g$ values found in the fitting of the UVES spectra to values estimated from the MIST isochrones (Fig. \ref{fig: logg comparison}). The $\log g$ estimation from the isochrones was done in the same manner as described in Sec. \ref{sec: param from spectra} for the stars observed with GIRAFFE. Based on the comparison of 62 UVES stars, the $\log g$ from the isochrones is on average $0.2\,$dex lower than from the spectra. More than half of the isochrone $\log g$ values differ less than $10\%$ from the spectral ones, and all but one are within $30\%$.

\centering
\includegraphics[width=8.8cm]{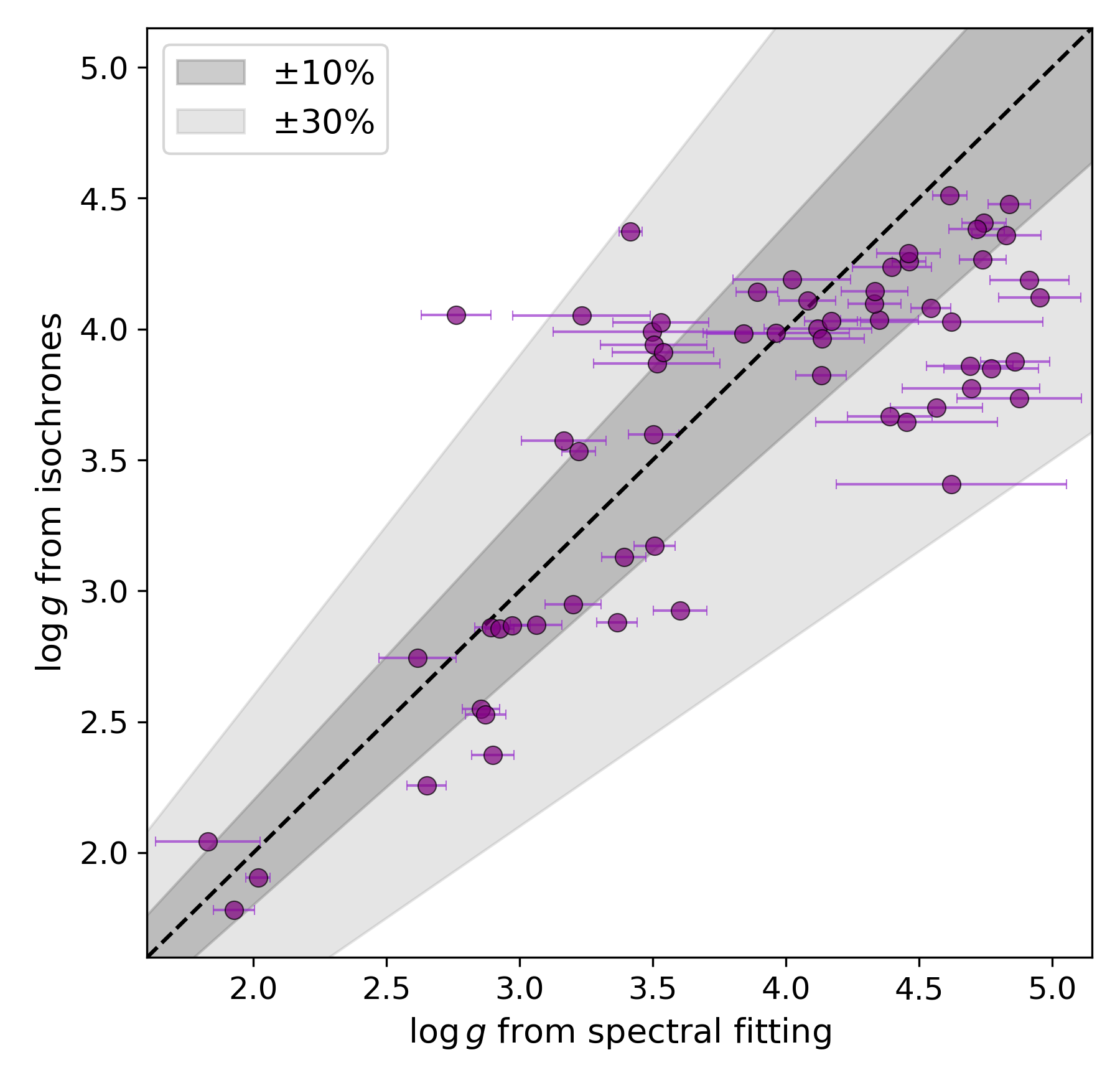}
\captionof{figure}{Parameter $\log g$ derived from the fitting of the UVES spectra compared to the $\log g$ estimated from the MIST isochrones.}
\label{fig: logg comparison}

\section{Parameter correlations} \label{appendix: correlations} 
\begin{justify}
We studied correlations between the parameters $A_{\mathrm{V}}$, $R_{\mathrm{V}}$, $\theta_{\mathrm{LD}}$, $\left[\mathrm{M}/\mathrm{H}\right]$, $T_{\mathrm{eff}}$, and $\log g$ (Fig. \ref{fig: correlations}) in the SED fitting (see Sec. \ref{sec: SEDs}). $A_{\mathrm{V}}$, $R_{\mathrm{V}}$, and $\theta_{\mathrm{LD}}$ were free parameters, while $\left[\mathrm{M}/\mathrm{H}\right]$, $T_{\mathrm{eff}}$, and $\log g$ were included as Gaussian priors based on the results from the previous steps. As expected, we see strong correlation between $T_{\mathrm{eff}}$ and $A_{\mathrm{V}}$, which was the main motivation to derive $T_{\mathrm{eff}}$ from spectra. There is also high positive correlation between the parameters $R_{\mathrm{V}}$ and $\theta_{\mathrm{LD}}$, which likely impacted their values resulting from the SED fits. On the contrary, $A_{\mathrm{V}}$ does not show significant correlation with $\theta_{\mathrm{LD}}$ and only moderate correlation with $R_{\mathrm{V}}$, which raises confidence on the derived field star and Cepheid extinction values.
\end{justify}

\includegraphics[width=8cm]{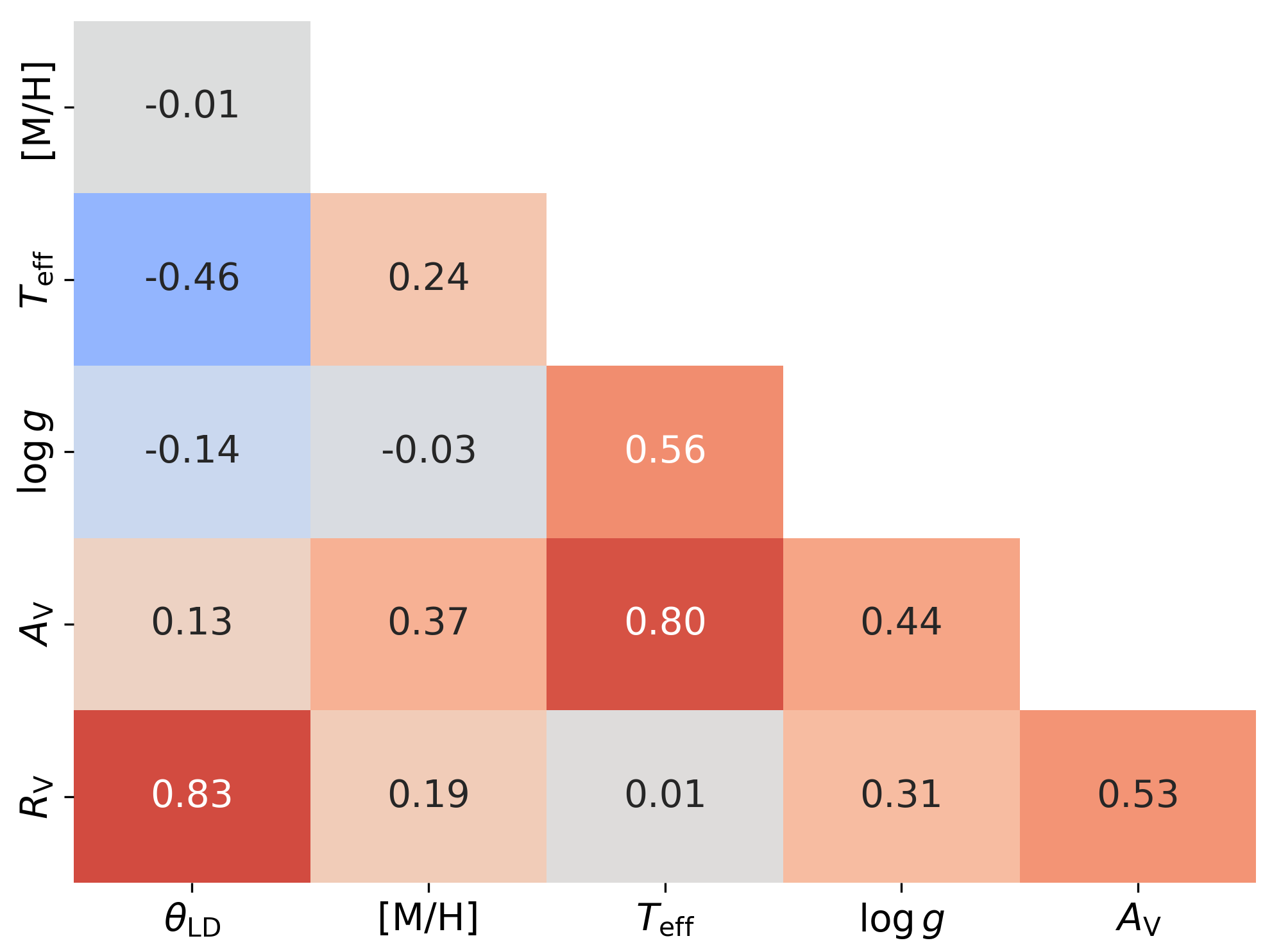}
\includegraphics[width=8cm]{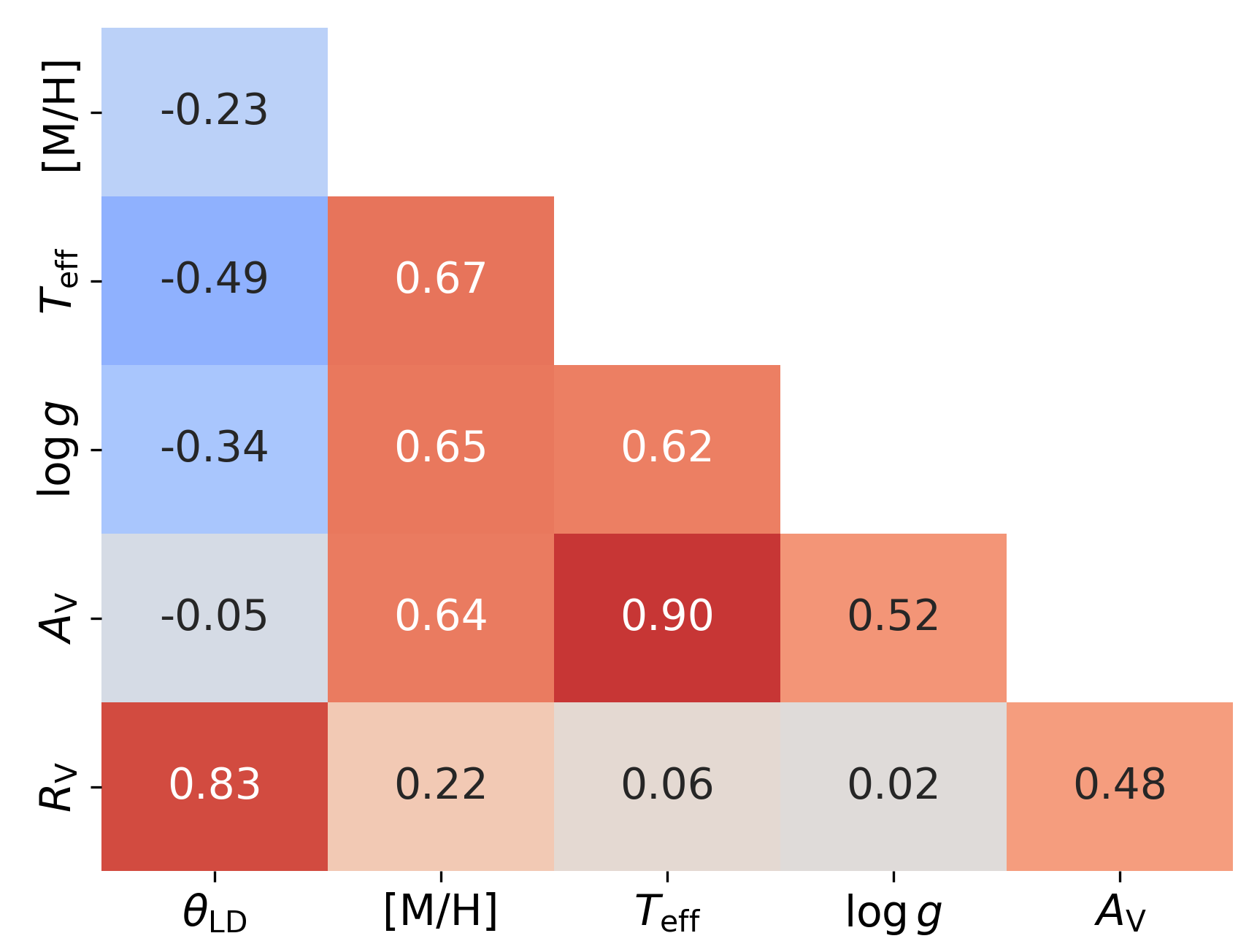}
\captionof{figure}{Triangle heatmaps showing the Pearson ($r$) correlation between parameters in the SED fitting. The maps are for the RS Pup field stars Gaia DR3 5546480122794409472 (upper) observed with GIRAFFE and Gaia DR3 5546476858619276672 (lower) observed with UVES.}
\label{fig: correlations}

\end{multicols}

\section{Comparison of the Cepheid extinction results to the literature} \label{appendix: histo comparison}

\centering
\includegraphics[width=18cm]{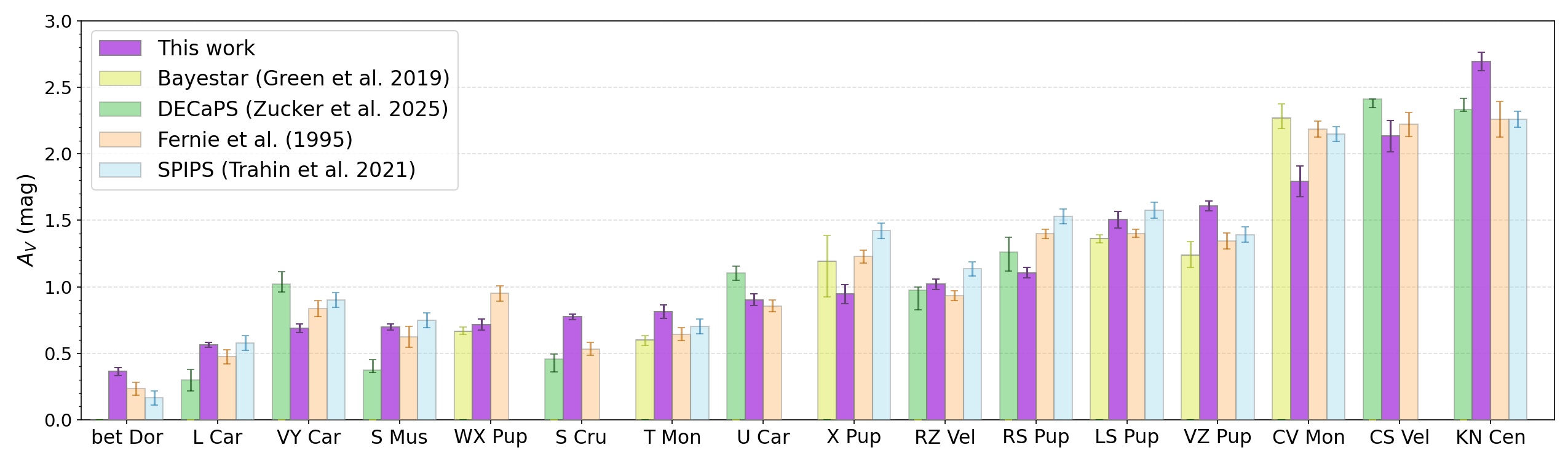}
\captionof{figure}{Comparison of our extinction results to literature values Cepheid by Cepheid. All of the literature values have been converted from $E_{\mathrm{B-V}}$ to $A_{\mathrm{V}}$, and the values from \citet{Fernie1995} and \citet{Trahin2021} have been rescaled as described in Sec. \ref{sec: literature comparison}.}
 \label{fig: histo_comparison}

\end{appendix}

\end{document}